\documentclass[3p, authoryear]{elsarticle}
\usepackage{amsmath}
\usepackage{amssymb}
\usepackage{multirow}
\usepackage{multicol}
\usepackage{dblfloatfix}
\usepackage{setspace}
\usepackage{lineno}
\usepackage{subfig}
\usepackage{lscape}
\usepackage{changepage}
\usepackage{psfrag}
\usepackage{color}
\usepackage{colortbl}
\usepackage{rotating}
\usepackage{tablefootnote}
\usepackage{arydshln}
\usepackage{gensymb}
\usepackage{url}
\usepackage{soul,xcolor}
\usepackage{caption}
\setstcolor{red}

\newcommand{\bA}{ {\boldsymbol A} }

\newcommand{\bI}{ {\boldsymbol I} }

\newcommand{\bK}{ {\boldsymbol K} }

\newcommand{\bs}{ {\boldsymbol s} }

\newcommand{\bu}{ {\boldsymbol u} }

\newcommand{\bv}{ {\boldsymbol v} }

\newcommand{\bw}{ {\boldsymbol w} }

\newcommand{\by}{ {\boldsymbol y} }

\newcommand{\bepsilon}{ {\boldsymbol \epsilon} }

\newcommand{\btheta}{ {\boldsymbol \theta} }

\newcommand{\bzero}{ {\boldsymbol 0} }

\newcommand{\norm}{ {\mathcal N} }

\definecolor{Gray}{gray}{0.9}
\newcolumntype{g}{>{\columncolor{Gray}}l}

\newcommand{\specialcell}[2][c]{
  \begin{tabular}[#1]{@{}c@{}}#2\end{tabular}}

\title{Geostatistical estimation of forest biomass in interior Alaska combining Landsat-derived tree cover, sampled airborne lidar and field observations}

\author[uw]{Chad Babcock\corref{cor1}}
\author[msu]{Andrew O. Finley}
\author[usfs]{Hans-Erik Andersen}
\author[usfs]{Robert Pattison}
\author[nasa]{Bruce D. Cook}
\author[nasa]{Douglas C. Morton}
\author[amer]{Michael Alonzo}
\author[nasa]{Ross Nelson}
\author[yale]{Timothy Gregoire}
\author[nmbu]{Liviu Ene}
\author[nmbu]{Terje Gobakken}
\author[nmbu]{Erik N{\ae}sset}

\cortext[cor1]{Corresponding author}

\address[uw]{School of Environmental and Forest Sciences, University of Washington, Seattle, WA}
\address[msu]{Forestry Department, Michigan State University, East Lansing, MI}
\address[usfs]{USDA Forest Service, Pacific Northwest Research Station, Seattle, WA}
\address[nasa]{Code 618, Biospheric Sciences Branch, NASA/Goddard Space Flight Center, Greenbelt, MD}
\address[amer]{Department of Environmental Science, American University, Washington, DC}
\address[yale]{Yale School of Forestry and Environmental Studies, Yale University, New Haven, CT}
\address[nmbu]{Faculty of Environmental Sciences and Natural Resource Management, Norwegian University of Life Sciences, \r{A}s, Norway}

\begin{document}

\begin{frontmatter}

\begin{abstract}
Lidar provides critical information on the three-dimensional structure of forests. However, collecting wall-to-wall laser altimetry data at regional and global scales is cost prohibitive. As a result, studies employing lidar for large area estimation typically collect data via strip sampling, leaving large swaths of the forest unmeasured by the instrument. The goal of this research was to develop and examine the performance of a coregionalization modeling approach for combining field measurements, strip samples of airborne lidar and Landsat-based remote sensing products to predict aboveground biomass (AGB) in interior Alaska's Tanana Valley. The proposed modeling strategy facilitates mapping of AGB density across the domain. Additionally, the coregionalization framework allows for estimation of total AGB for arbitrary areal units within the study area---a key advance to support diverse management objectives in interior Alaska. This research focuses on characterization of prediction uncertainty in the form of posterior predictive coverage intervals and standard deviations. Using the framework detailed here, it is possible to quantify estimation uncertainty for any spatial extent, ranging from point-level predictions of AGB density to estimates of AGB stocks for the full domain. The lidar-informed coregionalization models consistently outperformed their counterpart lidar-free models in terms of point-level predictive performance and total (mean) AGB precision. Additionally, including a Landsat-derived forest cover covariate further improved precision in regions with lower lidar sampling intensity. Findings also demonstrate that model-based approaches not explicitly accounting for residual spatial dependence can grossly underestimate uncertainty, resulting in falsely precise estimates of AGB. The inferential capabilities of AGB posterior predictive distribution (PPD) products extend beyond simply mapping AGB density. We show how PPD products can provide insight regarding drivers of AGB heterogeneity in boreal forests, including permafrost and fire, highlighting the range of potential applications for Bayesian geostatistical methods to integrate field, airborne and satellite data. 
\end{abstract}


\end{frontmatter}

\section{Introduction}
Coupling remote sensing data with field-based forest measurements via regression frameworks offers the potential to increase the precision of inventory estimates and provides a mechanism for mapping the spatial distribution of forest biophysical properties. A plethora of studies show strong relationships between lidar metrics and forest variables \citep{asner2009, babcock2013, finley2014, finley2017, lim2003, naesset2004, naesset2011}. These findings have spurred investment in collecting lidar data for large areas from aircraft and satellites alike. Of particular interest is the use of lidar to assist in the estimation of forest inventory parameters in high-latitude terrestrial ecosystems. From a carbon monitoring perspective, forests in boreal systems may contain large stores of aboveground biomass (AGB) and carbon, but the uncertainty associated with current estimates is extremely high \citep{bradshaw2015, pan2011}. Understanding that the taiga-tundra ecotone is one of the most vulnerable environmental systems to climate change and that its boreal forests can contribute substantially to the global carbon cycle, methods are needed to begin monitoring forest carbon stocks and fluxes for these systems \citep{gauthier2015, magnani2007, neigh2013}. 

Current approaches used by the United States Forest Service's (USFS) Forest Inventory and Analysis (FIA) program to quantify AGB and carbon stocks in temperate regions rely on extensive, spatially-balanced field plot probability samples to generate forest inventory estimates with acceptable levels of precision \citep{bechtold2005, woodall2015}. In vast remote landscapes, implementing the estimation techniques used by the FIA in the contiguous United States becomes prohibitively expensive due to the high cost of collecting field inventory data in difficult-to-access boreal regions, e.g., interior Alaska \citep{barrett2011}. A potential solution commonly put forward to reduce the expense of monitoring AGB in boreal forest systems is to augment sparse collections of field samples with remote sensing auxiliary data \citep{wulder2012}. Lidar-derived measures of forest structure tend to be highly correlated with AGB field observations and, thus, are prime candidates to supplement boreal field campaigns. Additionally, passive sensors such as Landsat can be used to derive remote sensing data products correlated with forest AGB \citep{kumar2015}. Methodologies leveraging relationships between field and lidar can potentially be further improved by incorporating Landsat-based products \citep{margolis2015, pflugmacher2014, powell2010, zheng2004}.

Here, we address two challenges encountered when attempting to estimate forest AGB for large areas using lidar coupled with other remote sensing information: 1) incomplete spatial coverage of remote sensing data; and 2) prediction uncertainty quantification. Incomplete spatial coverage is a common problem for studies using airborne or spaceborne lidar over sizable study domains \citep{andersen2011, bolton2013, nelson2010, nelson2004}. Model-based methodologies used to link field and lidar data to estimate and map AGB typically require laser altimetry information for the entire spatial domain of interest \citep{babcock2015, babcock2016, mcroberts2013}. The expansive nature of boreal systems, make wall-to-wall collections of airborne lidar data unrealistic. Further, future spaceborne lidar systems are not designed to procure complete coverage information. Rather, these campaigns will collect data for relatively narrow bands along the orbital tracts of the sensors' host satellite \citep{GEDI2014, ICESAT2015}. In order to glean any additional information provided by sampled remote sensing data in a statistically rigorous manner, estimation frameworks that can accommodate incomplete coverage auxiliary information are necessary. 

The second issue examined here is the problem of obtaining useful estimates of uncertainty about forest AGB stocks using model-based statistical procedures---necessary for decision making with imperfect predictions of forest AGB. In design-based estimation frameworks, error is assumed to arise from the sampling design, which can be appropriately characterized when plots are selected probabilistically \citep{cochran1977, thompson2002}. In model-based inference, error is attributed to the underlying process by which the response, e.g., AGB, is generated \citep{gregoire1998, verhoef2002}. Studies attempting to estimate means and totals for areal units using ancillary data within a model-based paradigm need to specify frameworks that reliably accommodate the structure of the data to be modeled. It can be the case that modelers who attempt to use model-based forest inventory estimation approaches posit potentially unrealistic assumptions about the distributional characteristics of model errors, such as independent and identically distributed (\emph{iid}) errors. In a spatial context, it is likely the field observations of AGB will be spatially autocorrelated. If the auxiliary information used in the model fails to fully account for the spatial dependence among field observations, model-based approximations of AGB uncertainty can be grossly underestimated \citep{cressie1993, griffith2005}.

Coregionalization models constructed within a Bayesian hierarchical framework offer a solution to both above-mentioned challenges \citep{gelfand2004}. This class of multivariate spatial regression  models is designed to predict multiple response variables simultaneously while leveraging spatial cross-correlation structures between error components of the responses. Further, the model can accommodate spatial misalignment, i.e., missing response variable measurements at some locations. If the lidar data is treated as an explanatory variable (used on the right-hand side of the model as in most lidar studies) predictions are only possible where lidar data is available. Within a coregionalization model, the lidar-derived metrics can be treated as additional response variables (moved to the left-hand side) and jointly predicted with the response of interest, e.g., AGB, across the entire landscape while explicitly modeling the spatially co-varying relationship among the predictions within and across locations \citep{finley2014b}. A coregionalization framework also allows for the inclusion of wall-to-wall covariates derived from satellite data to assist in the joint prediction of forest AGB and lidar information.

When multivariate coregionalization models are estimated using a Bayesian hierarchical approach, uncertainty occurring at all levels of the model can be propagated through to prediction and subsequent estimation of means and totals for areal units \citep{berliner1996, cressie2011, gelfand1990, hobbs2015}. Forms of multivariate spatial prediction models have been in existence since the 1960s, e.g., cokriging \citep{matheron1963}. These non-hierarchical implementations, however, struggle to effectively deal with uncertainty associated with spatial covariance parameters, e.g., spatial variances and decays \citep[Section 7.1.1]{diggle2007}. Due to increased computational efficiencies gained by ignoring uncertainty in spatial variability, `plug-in' spatial covariance parameters are used in cokriging interpolation routines available in popular GIS software packages. This limits their use for fully model-based predictive inference \citep{schelin2010}.

The development of inferential approaches for complex spatial prediction within a statistical framework is an active area of research. In a hierarchical modeling context, coregionalization frameworks can be constructed using random effects that arise from spatially correlated Gaussian processes and partition variability into spatial and non-spatial components \citep{banerjee2014,cressie2009}. When formulated as such, estimation approaches including Restricted Maximum Likelihood (REML) or Markov chain Monte Carlo (MCMC) become possible in frequentist and Bayesian paradigms of statistical model-based inference, respectively \citep{verhoef2004}. There are advantages to choosing a Bayesian hierarchical approach to inference over counterpart frequentist methods. Access to the full posterior predictive distribution (PPD), a by-product of Bayesian inference, allows for easy posterior summarization of means or totals with associated uncertainty for the full spatial domain in addition to any sub-domains that may be of interest--even under back-transformation \citep{stow2006}. Access to PPDs facilitate subsequent, i.e., post model-fitting, analysis to inform ecological or management objectives while accounting for prediction uncertainty. However, these increases in flexibility come with substantial increases in computational demand.

The aim of this study is to develop and examine the performance of a statistical modeling framework that can 1) incorporate partial coverage lidar data and wall-to-wall Landsat products to improve AGB density prediction; and 2) accommodate spatially structured variability unaccounted for by covariates, thereby allowing for more reliable model-based characterizations of uncertainty (e.g., uncertainty intervals with intended coverage) and improved prediction accuracy. We look to the Tanana Inventory Unit (TIU) in interior Alaska to explore the potential for the proposed coregionalization model to estimate forest AGB stocking by coupling spatially sparse field inventory, partial coverage lidar and Landsat-derived tree cover data products in boreal landscapes. The USFS and National Aeronautics and Space Administration (NASA) collaborated on the collection of field inventory data using an augmented FIA sampling design and flight-line samples of lidar data in 2014. Within this study region, four areal domains containing systematic samples of field and lidar data serve as study sites in this analysis. We use model comparison to identify strengths and limitations of candidate modeling approaches and information sources. These comparisons highlight tradeoffs associated with different estimation methods and benefits of coregionalization modeling for large-area estimation of AGB using sampled remote sensing data. We also demonstrate the potential of Bayesian spatial hierarchical models for generating small-area forest parameter estimates and analysis of variability in ecosystem structure via PPD summarization. At one site, generating AGB PPDs at a watershed scale offered insight regarding the drivers of biomass variability in interior Alaska, including permafrost and fire. We also compare uncertainty intervals generated using the Bayesian coregionalization model with classical cokriging to highlight advantages to fitting spatial models using Bayesian inference.

\section{Background}

\subsection{Using lidar and Landsat data fusion for large-area forest AGB mapping}

There is increased interest in combining active and passive remote sensing data sources, such as lidar and Landsat, for forest variable mapping over large spatial extents. Lidar is attractive for forest height, volume and AGB prediction because it is able to gather information relevant to the vertical structure of forests. Using spatially explicit height information from lidar can lead to more accurate forest variable maps than those based solely on passive remote sensing information provided by Landsat or aerial photography. \citet{ediriweera2014} explored the use of wall-to-wall lidar and Landsat data to map AGB and found that models incorporating both remote sensing sources performed better than using either alone. However, it can be expensive to collect wall-to-wall lidar useful for forest structure characterization over large areas, e.g., state- or country-level. Given this limitation, there has been increased research examining the integration of sampled lidar data, often along flight-lines or orbital tracts, with wall-to-wall readily available passive imagery to improve inventory mapping efforts. \citet{deo2017} explored multiple regression and Random Forest approaches fusing strip-sampled lidar and Landsat metrics to predict AGB and showed improved model fits when both remote sensing sources were used as covariates. Their models, however, could not yield a wall-to-wall map due to incomplete spatial coverage of the lidar-derived covariates. \citet{yavasli2016} examined a two-stage processing approach; first predicting field measured AGB at lidar sample locations, then regressing those lidar based AGB predictions on wall-to-wall Landsat imagery to map AGB outside the lidar sampled areas. This type of two-stage approach has the potential to improve point predictions based on field observations, lidar samples and wall-to-wall imagery. However, without a way to carry uncertainty across successive modeling stages, such approaches fail to provide realistic prediction uncertainty estimates in final map products. Without a well designed statistical framework able to effectively propagate error stemming from all processing steps through to prediction, multi-step interpolation routines can only rely on holdout validation metrics to assess uncertainty. Estimates of overall average error provided by such metrics can be inadequate when decision-makers attempt to implement policy or management initiatives using spatially explicit map products. Most proposed forest measurement, reporting and verification (MRV) systems aim to provide interpretable measures of spatially explicit uncertainty that account for all error sources using statistical uncertainty intervals. Such MRV systems include the United Nations Programme on Reducing Emissions from Deforestation and Forest Degradation (UN-REDD) and NASA's Carbon Monitoring System (CMS) \citep{CMS2010, REDD2009}.

\subsection{Design-based and model-assisted forest inventory leveraging remote sensing data}

Forest inventory estimation is typically conducted using design-based statistical principles, relying on probability samples of field plots to obtain estimates of inventory parameters, such as mean tree density, total growing stock volume or AGB. Most national forest inventory programs, including the USFS FIA, have been developed with a design-based approach to uncertainty quantification in mind. The advent of readily available remote sensing information has led to new developments in design-based estimation approaches incorporating models to improve estimation. Concerning the use of incomplete spatial coverage lidar, studies employing model-assisted estimators within a design-based inferential paradigm have been proposed that assume multi-phase data collection schemes \citep{gregoire2011, saarela2015, gregiore2016}. In these studies, a model is employed to relate field and remote sensing data and differences are analyzed using design-based variance estimators to approximate error associated with mean or total AGB estimates. When attempting to use remote sensing auxiliary data to improve AGB estimates in a design-based paradigm, one only requires remote sensing data to cover the plot locations, which can be accomplished by collecting flight-line strips of auxiliary information \citep{nelson2012}. These model-assisted approaches show promise but there are some significant limitations to their practical use in remote sensing-aided forest inventory. Design-based model-assisted approaches can be insufficient due to the restrictive sampling designs they require. We identify four shortcomings of design-based inference in reference to using wall-to-wall and/or sampled remote sensing data for forest inventory below.
\begin{enumerate}
\item Any model-assisted estimator depends on plots and sampled remote sensing data being established as probability samples, which in practice, is not easy to implement \citep[Chap. 8-9]{sarndal1992}. Although technically a probabilistic sampling approach, the systematic sampling designs which are typically conducted (or assumed) offer no means to derive a variance estimator. Rather variance estimators for other sampling designs, such as simple random sampling, are used to approximate systematic sampling variance with the understanding that uncertainty will likely be overestimated \citep{cochran1946}. The simulation study comparing different model-assisted estimation approaches presented in \citet{ene2012} exemplifies this phenomenon, showing that in the case of two-phase systematic sampling, uncertainty was over estimated by a factor of four. 
\item Multi-phase designs require the primary sampling units (field observations) to be a subsample of the secondary sampling units (lidar strips/grid cells). This renders any AGB field samples missed by lidar flight-lines or orbital tracks useless for estimation purposes. This restriction becomes particularly problematic when combining national-scale forest inventory plots and space-based lidar data to improve AGB estimation. This can result in estimates relying on field observations alone (e.g., FIA data) being more precise than those incorporating sampled lidar information simply because the direct estimator uses all available field samples, i.e., larger sample size. 
\item Model-assisted approaches can provide unstable small-area or post-stratified estimates of forest variables when excessively low field sample sizes are encountered in sub-domains \citep{breidenbach2012}. Further, design-based confidence intervals require assumptions about asymptotic normality. In small sample size situations, this assumption can be unrealistic. Also, design-based approaches offer no sound methodology for estimating the variable of interest with uncertainty in sub-domains where no field samples were collected. \citet{pfeffermann2013} provides further general discussion on the shortcomings of design-based estimation strategies concerning small-area estimation.
\item Model-assisted approaches leveraging sampled lidar data provide no mechanism for wall-to-wall mapping of the variable of interest with uncertainty. The ability to generate mean or total estimates of AGB alongside maps of AGB density, both with associated measures of error, can be advantageous for forest management and decision-making.
\end{enumerate}

\subsection{Geostatistics to quantify error in forest inventory and mapping with remote sensing data}

Geostatistical techniques, such as kriging and cokriging (in all their variants), were originally developed in the early 1960s as statistical geology tools to estimate underground ore reserves based on nearby field samples \citep{matheron1963, cressie1990}. Created specifically to obtain optimal interpolation at unknown locations (points or regions) based on ground-truth data, e.g., best linear unbiased predictions (BLUP), geostatistical approaches are uniquely relevant to predictive mapping problems throughout environmental remote sensing, including forest inventory applications. \citet{hudak2002} analyzed five lidar and Landsat-based modeling approaches to predict forest canopy height and showed that regressions incorporating spatial information using geostatistical techniques were better predictors than their non-spatial counterparts. \citet{tsui2013} examined several spatial modeling approaches, including regression-kriging and cokriging, to interpolate lidar-based predictions of AGB between simulated flight-line strips leveraging wall-to-wall space-based radar. They found that geostatistical modeling incorporating wall-to-wall covariates, e.g., regression-kriging, was helpful for extrapolating between lidar strip samples. 

The geostatistical studies mentioned above, along with many others found in the remote sensing and forest inventory literature, focus only on optimal interpolation for accurate mapping, disregarding the potential for spatial modeling to characterize uncertainty associated with predictions \citep{meng2009, mutanga2006}. For good reason, kriging and cokriging standard errors are often considered underestimates of uncertainty and therefore ignored \citep[Section 7]{diggle2007}. This is due to the use of `plug-in' semi-variogram (or autocovariance) model parameters within the kriging/cokriging prediction routine. By interpreting classical kriging standard errors and subsequent confidence intervals as statistical measures of uncertainty, the researcher assumes the estimated spatial covariance parameters, e.g., nugget, partial sill and range, are known without error. In typical applications this assumption is far from true. Spatial covariance parameters based on semi- and cross-variogram models depend on many factors, including how observation pairs are binned and variogram model selected to fit to the empirical variogram points, e.g., exponential, spherical or Mat\'{e}rn. In addition, the method used to estimate semi- and cross-variogram model parameters, whether by maximum likelihood, eye-fitting or others, can lead to very different parameter estimates and all ignore the inherent variability around the resulting point estimates. Hierarchical Bayesian approaches to spatial model fitting allow researchers to relax these restrictive assumptions, leading to uncertainty characterizations that can be more safely interpreted as statistical measures of error. By allowing modelers to specify appropriately vague priors for spatial covariance parameters within a hierarchical framework, Bayesian estimation naturally propagates uncertainty into the posterior (predictive) distribution. Resulting uncertainty characterizations based on the posterior distribution of predictions therefore account for error about the spatial covariance parameters, especially when vague prior distributions are used during fitting \citep{gelman2013}. Bayesian hierarchical modeling approaches are also easily extendable to multivariate response settings with incomplete observations, e.g., field AGB and airborne lidar samples \citep{gelfand2004}.

\section{Methods}

\subsection{Study sites}
\begin{figure}[!b]
  \centering
  \includegraphics[width = \textwidth]{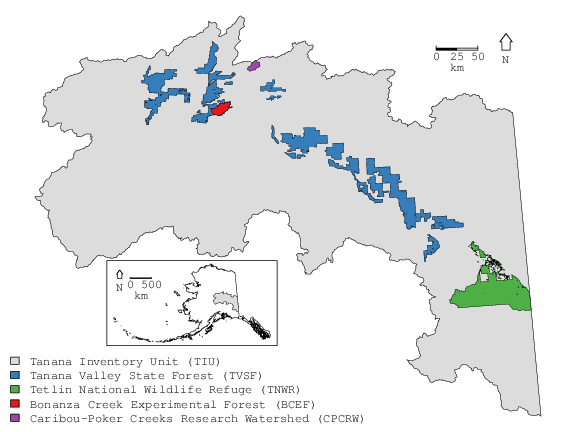}
  \caption{Map showing locations of the four study sites within the Tanana Inventory Unit (TIU).}\label{map}
\end{figure}

The four study sites explored in this analysis all fall within the TIU in interior Alaska (Fig.~\ref{map}). The Tanana Valley State Forest (TVSF) is a 730\,000 hectare (ha) tract of predominately boreal forest stretching along the Tanana river basin. Nearly 90 percent of the TVSF is considered forested and close to 50 percent of all productive forestland in the TIU is contained within the boundary of the TVSF \citep{TVSF2016}. Tree species typical of taiga forest can be found throughout, including white spruce (\emph{Picea glauca}), black spruce (\emph{Picea mariana}), tamarack (\emph{Larix laricina}), quaking aspen (\emph{Populus tremuloides}) and balsam poplar (\emph{Populus balsamifera}).
 
Tetlin National Wildlife Refuge (TNWR) is nearly 300\,000 ha in size with lowland areas characterized by extensive wetlands and poorly drained soils. Wet upland sites are home to black spruce forests whereas drier landscapes favor white spruce. Deciduous species including quaking aspen, paper birch (\emph{Betula papyrifera}) and balsam poplar persist on well-drained south-facing slopes. Shrub vegetation consisting of willow (\emph{Salix spp.}), alder (\emph{Alnus spp.}) and dwarf birch (\emph{Betula spp.}) can be found in lowland areas around water bodies \citep{TNWR2016}.

Bonanza Creek Experimental Forest (BCEF) is a Long-Term Ecological Research (LTER) site within the TVSF, consisting of vegetation and landforms typical of interior Alaska. The BCEF domain delineated for this study is 21\,000 ha and includes a section of the Tanana River floodplain along the southeastern border. The BCEF is a mixture of forest and non-forest vegetation compositions featuring white spruce, black spruce, tamarack, quaking aspen and balsam poplar trees mixed with willow and alder shrubland species \citep{BCEF2016}.

Caribou-Poker Creeks Research Watershed (CPCRW) is an intensively studied basin reserved for hydrological and ecological research \citep{CPCRW2016}. CPCRW is approximately 10\,600 ha in size and divided into 11 watershed units. Many research initiatives at CPCRW involve vegetation and hydrology comparisons among watershed units \citep{amatya2016, rinehart2015, tanaka2016}. Upland areas are dominated by paper birch and aspen on south-facing slopes. North-facing slopes are largely occupied by black spruce. Patches of alder exist in the understory. Lowland sites are composed of moss and dwarf shrubs. CPCRW has a rich source of associated mapped data products including permafrost polygon layers and fire maps \citep{chapin2006, rieger1972, verbyla2011}. In 2004, the Boundary Fire in interior Alaska burned a significant portion of CPCRW, predominantly along the southeastern border of the research area \citep[Fig.~1]{hollingsworth2013}. 

At all sites, AGB density was not predicted over \emph{Water} and \emph{Barren Lands} grid cells classified by the National Land Cover Database \citep{homer2015}. These areas were not considered to be part of the study domains. 

\subsection{Field measurements}
At TVSF and TNWR, a spatially-balanced systematic design was implemented using a tessellation of hexagons covering the study areas. The hexagons were approximately 12\,141 ha (30\,000 acres) in size. These polygons are five times larger than the hexagons used to establish plots in the continental United States. In the summer of 2014, FIA field crews established standard FIA plots at the center of each hexagon complete with four subplots; one at the midpoint and three additional subplots extended radially approximately 36 meters (m) at $0^\circ$, $120^\circ$ and $240^\circ$. Each subplot has an approximate 7.3 m radius and an area of 168.11 m$^2$. Field measurements were taken using an augmented FIA inventory design \citep{bechtold2005}. A notable change to the typical protocol included the addition of a second micro-plot to be inventoried in each subplot. Adding a second micro-plot helped to ensure that a sufficient number of small diameter trees were tallied (between 2.5 and 12.7 centimeter (cm) diameter at breast height) \citep{andersen2011}. \citet{pattisonInPrep} provide detailed field protocols for the TIU inventory pilot project. The same sampling protocol was used at BCEF and CPCRW, although at a greatly increased sampling intensity. Field plots at BCEF and CPCRW were each inventoried once in the summer months of 2011, 2012 or 2014. We recognize the mismatch between field plot measurement and lidar acquisition years for some plots at BCEF and CPCRW is not ideal, but given that boreal forests in this region are slow growing, we argue that this level of temporal misalignment will have a negligible affect on subsequent results. The total number of subplots measured at each site was 263, 123, 292 and 149 at TVSF, TNWR, BCEF and CPCRW, respectively (note that some subplots at each plot were not measured for various logistical reasons or fell outside the study area boundary making the total number of subplots not wholly divisible by four). AGB for individual trees on subplots were tabulated using the Component Ratio Method described in \citet{woodall2015}. AGB for all trees on a subplot with breast height diameters 2.5 cm and larger were scaled to megagrams per ha (Mg/ha) and summed to obtain subplot-level AGB density. Prior to model-fitting, a square-root transformation was applied to the subplot-level AGB densities to better approximate a Gaussian error distribution and ensure positive support following back-transformation of predicted values. For this analysis each inventory subplot was treated as a distinct observation. To ease explanation, the term \emph{plot} will be used in subsequent sections to refer to the individual FIA subplots described above.

\subsection{Remote sensing data}\label{rsdata}
Lidar data were collected using a flight-line strip sampling approach with NASA Goddard's LiDAR, Hyperspectral and Thermal (G-LiHT) airborne imager in the summer of 2014 \citep{cook2013}. G-LiHT is a portable multi-sensor system that is used to collect fine spatial-scale image data of canopy structure, optical properties, and surface temperatures. G-LiHT's on-board laser altimeter (VQ-480, Riegl Laser Measurement Systems, Horn, Austria) provides an effective measurement rate of up to 150 kilohertz along a 60\degree\,swath perpendicular to the flight direction using a $1\,550$ nanometer laser. At a nominal flying altitude of 335 m, laser pulse footprints were $\approx 10$ cm diameter and sampling density was $\approx 6$ laser pulses m$^{-2}$. Pulse densities $>4$ m$^{-2}$ are able to more accurately characterize complex terrain and vertical distribution of canopy elements in dense stands, and tree height metrics (e.g., 90$^{\text{th}}$ percentile height used here) are largely unaffected by pulse densities above 1 pulse m$^{-2}$ \citep{jakubowski2013,white2013}. G-LiHT's lidar is capable of producing up to eight returns per pulse. Point cloud information was summarized to a 13 x 13 m grid cell size (grid cell area equal to 169 m$^2$) to approximate field plot areas. Over each grid cell, percentile heights were calculated at 10 percent intervals ranging from 10 percent to 100 percent. Maximum height (100$^{\text{th}}$ percentile height) relative densities were also calculated at 10 equal width intervals. Additional metrics including point cloud skewness and kurtosis, among others, were calculated for each grid cell as well. Identical lidar metrics were obtained using point clouds extracted over each field plot. Exploratory regression model fits indicated that 90$^{\text{th}}$ percentile height alone accounted for substantial amounts of variability in square-root transformed AGB at all sites ($0.78 \le R^2 \le 0.87$). Due to the high correspondence between 90$^{\text{th}}$ percentile height and AGB, it was decided that only this metric will be considered for subsequent modeling efforts. G-LiHT data for the study area are available online at \url{https://gliht.gsfc.nasa.gov}.

To reduce computational demand associated with fitting the proposed models, the gridded lidar covariate sets were subsampled. At the four study sites, approximately 0.5 percent of the original lidar grid cells were randomly selected. The lidar subset grid cells were combined with the 90$^{\text{th}}$ percentile height values calculated using lidar point clouds over the field plots to construct the lidar metric sets for subsequent model-fitting. Specifically, 1\,216 of 199\,045, 813 of 132\,635, 2\,011 of 491\,621 and 7\,069 of 1\,374\,118 lidar observations were used for model-fitting at BCEF, CPCRW, TNWR and TVSF, respectively. Increases in computational power in the future will reduce the need for substantial thinning of the lidar metric set, allowing for the models tested here to be fitted using more observations. Also, see Section \ref{discussion} for discussion about future modeling research avenues that may lead to increased computational efficiencies, thereby alleviating the need to substantially thin sampled remote sensing datasets to implement the modeling frameworks proposed here.

We examined several Landsat-derived products to evaluate the potential for additional predictive gains over the use of lidar data alone.  Candidate data layers with wall-to-wall coverage over the study area included percent tree cover for 2010 \citep{hansen2013}, individual reflectance bands from Landsat 8 Operational Land Imager (OLI) composite surface reflectance products for 2014, and vegetation indices (e.g., Tasseled Cap values and Normalized Difference Vegetation Index, NDVI) derived from the 2014 Landsat composites. The 2010 percent tree cover data product exhibited the strongest relationship with field observations of AGB. Further, including 2014 Landsat composite bands and indices in addition to 2010 percent tree cover in regressions did not result in appreciable gains in fit performance. For these reasons, only the 2010 percent tree cover metric was considered for this analysis. $R^2$ values for preliminary models relating square-root transformed AGB density and percent tree cover ranged between $0.25$ and $0.55$ for the study areas.

\subsection{Model Overview}
\begin{figure}
  \centering
  \includegraphics[width=\textwidth]{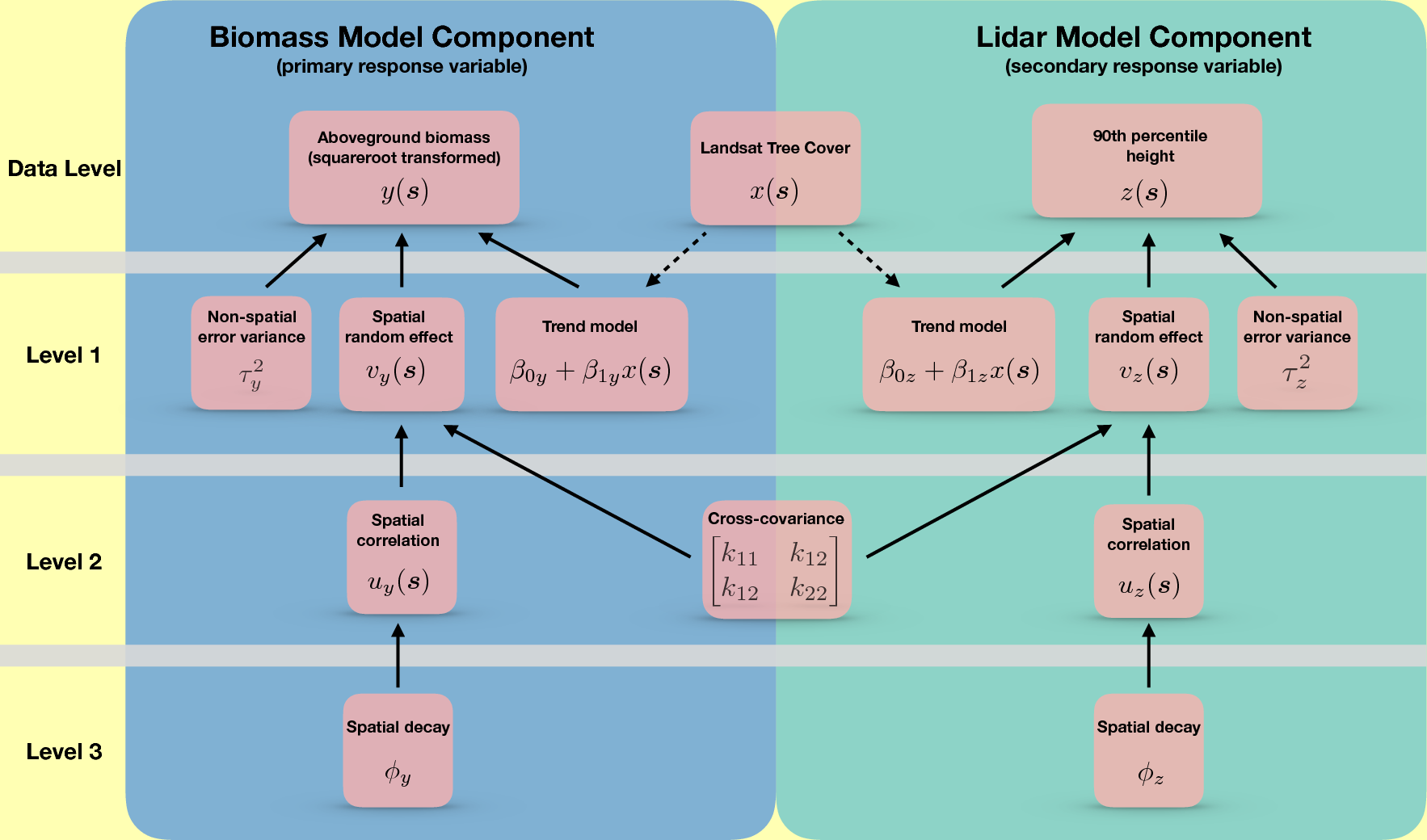}
  \caption{Overview schematic of the full hierarchical model, labeled as \emph{Coregionalization$+$Tree Cover} in the results tables. Solid black arrows indicate stochastic linkages for model parameterization, and dashed black arrows show deterministic links based on input data.  To fully specify this framework as a Bayesian hierarchical model, all parameters stochastically linked to other parameters or observations need to be assigned probability distributions. The blue rectangle contains all model elements (pink rectangles) associated with the aboveground biomass component and the green rectangle contains all model elements associated with the lidar component of the full model. The Landsat tree cover data and cross-covariance matrix belong to both model components, providing the link between the primary and secondary responses. The levels of the hierarchy are delineated with gray bars highlighting the multi-level structure of the model.}\label{dag}
\end{figure}

\subsubsection{Coregionalization model explanation}
Fig.~\ref{dag} is a graphical depiction of the full model tested in this analysis, labeled as \emph{Coregionalization$+$Tree Cover} in the results tables. The other models tested can be viewed as \emph{special cases} of the full model, formed by setting different model terms to zero. The goal of this section is to provide a high-level description of the model framework used in this study. Further details concerning each of the individual tested models are provided in Section \ref{candidates} and references therein. We point interested readers to \citet{hobbs2015} for an accessible overview of Bayesian modeling principles in general and \citet{banerjee2014} for specific discussion about the models tested in this analysis---including the coregionalization geostatistical framework.

Fig.~\ref{dag} shows distinct, but connected, components for each response variable---AGB (shown in blue) and lidar derived 90$^{\text{th}}$ percentile height (shown in green)---highlighting that this is a joint modeling framework. Level 1 of the hierarchical framework contains three elements for each response. The trend model element for both response components is a simple linear regression with an intercept and slope parameter associated with the Landsat derived tree cover data $x(\bs)$. We define $\bs$ to be a vector of geographic coordinates, e.g., easting and northing, in the spatial domain $\mathcal{D}$. In addition to the trend model, each component contains a spatial random effect ($v_y(\bs)$ and $v_z(\bs)$) designed to model any spatially structured variability left after accounting for the trend model. The variance parameters ($\tau_y^2$ and $\tau_z^2$) account for any uncorrelated error once the trend models and spatial random effects are considered. 

Level 2 of the hierarchical model contains the makings of the Level 1 spatial random effects, i.e., the $u_y(\bs)$ and $u_z(\bs)$ spatial correlation effects and a shared cross-covariance matrix $\bK$. The cross-covariance matrix parameters estimate the spatial variability for each spatial random effect ($k_{11}$ and $k_{22}$) and also their covariance $k_{12}$. The covariance parameter, $k_{12}$, is the link between the AGB and 90$^{\text{th}}$ percentile height model components and is informed by observations of spatially coinciding responses (i.e., locations where both AGB and lidar derived 90$^{\text{th}}$ are observed). For our setting, estimates of $k_{12}$ provide two benefits. First, for locations where neither response is observed, the joint prediction of AGB and lidar 90$^{\text{th}}$ percentile height will maintain their observed covariance. Second, in spatial misalignment settings the value of the observed response will inform the prediction of the missing response, e.g., prediction of AGB will be improved for locations where 90$^{\text{th}}$ percentile height is observed along the lidar transects. 

In addition to informing prediction where one or both of the responses are missing (via the $\bK$) the proposed model informs prediction by pooling information from spatially nearby observations through a spatial correlation function which is informed by a set of spatial correlation parameters that reside in Level 3. As formalized in the subsequent section, these parameters control the geographic range of spatial dependence exhibited by the random effects. 

\subsubsection{Model distributional assumptions and fitting}
Recall, our goal is to produce spatially explicit predictions of AGB density and estimates of total (mean) AGB, both with quantifiable measures of uncertainty that reflect the fact that none of the estimated parameters are known without error. Such error quantification requires a probabilistic modeling framework and some parameter distributional assumptions. Fig.~\ref{dag} identifies the parameters to be estimated and the hierarchal dependence of the model components, but like most graphical depictions, the distributional assumptions are omitted \citep{lunn2012}. We elect to fit the model described in Fig.~\ref{dag} using Bayesian model-based inference to handle the process of propagating uncertainty stemming from all stages of the hierarchy through to prediction and subsequent estimation of total (mean) AGB.

Beginning with the Data Level, we assume that square-root transformed AGB, $y(\bs)$, arises from a normal distribution with a mean equal to the trend model plus a spatial random effect, i.e., $\beta_{0y} + \beta_{1y}x(\bs) + v_y(\bs)$, and a variance parameter, $\tau_y^2$. Similarly, we assume lidar derived 90$^{\text{th}}$ percentile height, $z(\bs)$, to be distributed normally with a mean equal to the trend model plus a spatial random effect, i.e., $\beta_{0z} + \beta_{1z}x(\bs) + v_z(\bs)$, and a variance parameter, $\tau_z^2$.

In Level 1 of the hierarchy, we see the trend model intercept and slope in addition to the spatial random effect are unknown for both components. Moreover, the variance parameters, $\tau_y^2$ and $\tau_z^2$, also need to be estimated. To estimate these parameters within a Bayesian statistical paradigm, we need to set prior distributions for them. Further, these prior distributions should reflect the researchers belief about what these parameters are before observing the data used to fit the model. In our case, we have no prior knowledge about the value of the intercept and slope parameters so we establish prior distributions to be normal with a mean equal to 0 and a variance of $10\,000$, i.e., $\norm(0,10\,000)$. The large variance for these priors makes them sufficiently vague, i.e., this choice of prior should not influence posterior inference beyond the observed data. We also have no prior understanding about the true values for the $\tau_y^2$ and $\tau_z^2$ variance parameters aside from knowing they are non-negative. For the $\tau_y^2$ and $\tau_z^2$ parameters we assume an Inverse-Gamma ($IG$) prior with a mean equal to a reasonable starting value and an infinite variance. The infinite variance setting establishes these as vague priors. The spatial random effects in Level 1, $\bv(\bs) = (v_y(\bs), v_z(\bs))'$, are set equal to $\bA\bu(\bs)$, where $\bu(\bs) = (u_y(\bs), u_z(\bs))'$ and $\bA$ is the square-root of the cross-covariance matrix $\bK$. This is a deterministic link, meaning that we do not need to specify a distribution for $\bv(\bs)$ directly. Rather we specify distributions for the unknown parts of $\bv(\bs)$, namely $u_y(\bs)$, $u_z(\bs)$ and $\bK$.

In Level 2, both the $u_y(\bs)$ and $u_z(\bs)$ random effects are assumed to be distributed according to a \emph{Gaussian process} ($GP$) with a mean equal to zero and a variance defined by a spatial correlation function. The correlation functions chosen for this analysis were exponential decay functions which depend on separation distances between points in the domain and spatial decay parameters. This class of correlation function is very useful for modeling spatially autocorrelated variability. When points in the domain are close together, the correlation is near one. The further the two points are from each other, the closer the correlation is to zero. The rate at which the functions approach zero depends on the spatial decay parameters (the Level 3, $\phi_y$ and $\phi_z$ parameters). The $\bK$ matrix controls the variance of each of the random effects and also the covariance between them. Again, $\bK$ is unknown so we set a prior for it to be an Inverse-Wishart ($IW$) distribution. The $IW$ distribution is a multivariate extension of an $IG$ distribution that is useful for setting priors for variance-covariance matrices. The $IW$ distribution has two parameters that define its shape. The first is a degrees of freedom parameter which we set to two because we are modeling a two-variable (bivariate) variance-covariance matrix. The second parameter for the $IW$ prior is a variance-covariance matrix that we set to $.1\bI_2$, where $\bI_2$ is a 2x2 identity matrix. This is a vague prior specification. Without standardization, the $k_{12}$ covariance parameter (an element of the variance-covariance matrix $\bK$) can be difficult to interpret. Because of this, we present a correlation estimate between the spatial random effects in the results tables labeled $cor(u_y(\bs),u_z(\bs))$ which is a simple transformation of the estimated cross-covariance matrix.

Since the Level 3 spatial decay parameters, $\phi_y$ and $\phi_z$, need to be estimated, sensible priors need to be set for them. We set uniform prior distributions for $\phi_y$ and $\phi_z$ with a lower bound near zero and large upper bound in effort to make these priors as vague as possible. Spatial decay parameters are difficult to interpret directly, so we present \emph{effective range}'s in the results tables ($er_y$ and $er_z$) which are simply transformations of $\phi_y$ and $\phi_z$ to distance units (km). 

Now that we have defined the likelihood, random effect and prior distributions for all unknown parameters we can construct a sampling algorithm to draw from the posterior distribution of our model parameters. For this analysis we used the \texttt{spBayes} package written for the \texttt{R} statistical computing environment which contains a class of functions designed to fit models as defined in Fig.~\ref{dag}, namely, the \texttt{spMisalign} function.

\subsection{Candidate models}\label{candidates}
To assess the utility of coupling plot-level AGB measurements with sampled airborne lidar and the 2010 Landsat-based tree cover index developed by \citet{hansen2013}, six candidate models were compared. Candidate models were evaluated on fit to observed data, prediction performance and total (mean) AGB estimation precision. The following sections provide the statistical modeling details required to implement the six candidate models. The mathematical notation used in the following sections closely mirrors that of \citet{banerjee2014}.  

\subsubsection{Null}
The \emph{Null} model is designed to be a baseline regression where no information beyond plot measurements is considered. The \emph{Null} model is
\begin{equation}
y(\bs) = \beta_{0y} + \epsilon_y(\bs),\label{eq:null}
\end{equation}
where $y(\bs)$ is a square-root transformed field-based measurement of AGB at location $\bs$ in spatial domain $\mathcal{D}$. $\beta_{0y}$ is an intercept parameter to be estimated. Since the \emph{Null} model contains no additional regression parameters, the estimated value for $\beta_{0y}$ should approximate the overall mean of square-root transformed AGB from the field plots, i.e., $\sum_{i=1}^ny(\bs_i)/n$, where $n$ is the total number of field plots measured. The error term $\epsilon_y(\bs)$ captures any departure from the overall mean at site $\bs$. Imposing a distributional assumption on $\bepsilon_y \overset{iid}{\sim} \mathcal{N}(\bzero,\tau_y^2\bI)$---where $\bepsilon_y = (\epsilon_y(\bs_1), \epsilon_y(\bs_2),..., \epsilon_y(\bs_n))^\prime$, $\tau_y^2$ is a variance parameter to be estimated and $\bI$ is an $n$ x $n$ identity matrix---allows for model-based statistical inference concerning model parameters and predictions. Any statistical inference, e.g., credible intervals, posterior standard deviations, etc., evaluated for model parameters or predictions can be considered reliable if the distributional assumptions about $\bepsilon_y$ are not seriously violated.

\subsubsection{Tree Cover}
The \emph{Tree Cover} model is
\begin{equation}
y(\bs) = \beta_{0y} + \beta_{1y}x(\bs) + \epsilon_y(\bs),\label{eq:treecover}
\end{equation}
where $y(\bs)$ and $\epsilon_y(\bs)$ are defined as in model (\ref{eq:null}). The intercept, $\beta_{0y}$, and regression slope parameter, $\beta_{1y}$, together describe the linear relationship between the Landsat-based tree cover product, $x(\bs)$, and $y(\bs)$.

\subsubsection{Spatial}
Incorporating a spatial random effect into model (\ref{eq:null}) establishes the \emph{Spatial} model and is
\begin{equation}
y(\bs) = \beta_{0y} + w_y(\bs) + \epsilon_y(\bs),\label{eq:spatial}
\end{equation}
where $y(\bs)$, $\beta_{0y}$ and $\epsilon_y(\bs)$ are defined as in model (\ref{eq:null}). Here, $w_y(\bs)$ is modeled as a \emph{Gaussian process} ($GP$) with a zero mean and a spatial covariance function that captures the covariance between any pair of locations $\bs$ and $\bs^\ast$, i.e., $w_y(\bs) \sim GP(0, C(\bs,\bs^\ast;\btheta_y))$. We specify $C(\bs,\bs^\ast;\btheta_y) = \sigma^2_y\rho(\bs,\bs^\ast;\phi_y)$ where $\rho(\bs,\bs^\ast;\phi_y)$ is a valid spatial correlation function and $\btheta_y = \{\sigma^2_y,\phi_y\}$, where $\phi_y$ is a correlation decay parameter and $\sigma_y^2 = Var(\bw_y)$. For this analysis, we assume an exponential correlation function, i.e., $\rho(||\bs - \bs^\ast||;\phi_y) = \exp(-\phi_y||\bs - \bs^\ast||)$, where $||\bs - \bs^\ast||$ is the Euclidean distance between locations $\bs$ and $\bs^\ast$ in kilometers (km). This specification for the spatial correlation function requires the modeler to assume the spatially structured error variability to be stationary and isotropic, meaning that spatial variability left after accounting for the trend does not depend on location or direction. This is true for the spatial random effects defined in subsequent sections as well. To ease interpretation of the $\phi_y$ estimates, corresponding effective range estimates, labeled $er_y$, are presented in the results tables. We define $er_y$ as the distance (km) where the spatial correlation between locations drops to 0.05. See \citet{gelfand2004} for specifics on how to calculate effective spatial ranges using coregionalization models.

\subsubsection{Spatial$+$Tree Cover}
Introducing a spatial random effect in model (\ref{eq:treecover}) defines the \emph{Spatial$+$Tree Cover} candidate written as 
\begin{equation}
y(\bs) = \beta_{0y} + \beta_{1y}x(\bs) + w_y(\bs) + \epsilon_y(\bs),\label{eq:spatial+treecover}
\end{equation}
with all terms defined previously.

\subsubsection{Coregionalization}\label{sec:coreg}
A bivariate coregionalization model for the joint prediction of square-root transformed AGB and lidar-derived 90$^{\text{th}}$ percentile height is
\begin{equation}
\begin{bmatrix}y(\bs)\\z(\bs)\end{bmatrix} = \begin{bmatrix}\beta_{0y}\\\beta_{0z}\end{bmatrix} +\bA\begin{bmatrix}u_y(\bs)\\u_z(\bs)\end{bmatrix} + \begin{bmatrix}\epsilon_y(\bs)\\\epsilon_z(\bs)\end{bmatrix},\label{eq:coreg}
\end{equation}
where $y(\bs), \beta_{0y}$ and $\epsilon_y(\bs)$ are defined previously. $z(\bs)$ is 90$^{\text{th}}$ percentile height recorded at location $\bs$ and, because no other covariates appear in the sub-model for $z(\bs)$, $\beta_{0z}$ approximates the overall mean of the sampled 90$^{\text{th}}$ percentile heights. Here, $u_y(\bs)$ and $u_z(\bs)$ are modeled as $GP$s with zero means and spatial correlation functions (rather than covariance functions as with $w_y(\bs)$ in \emph{Spatial} and \emph{Spatial$+$Tree Cover} models), i.e., $u_y(\bs) \sim GP(0, \rho(\bs,\bs^\ast;\phi_y))$ and $u_z(\bs) \sim GP(0, \rho(\bs,\bs^\ast;\phi_z))$, where $\phi_y$ and $\phi_z$ are spatial decay parameters. We again assume an exponential correlation function for both random effects, i.e., $\rho(||\bs - \bs^\ast||;\phi_y) = \exp(-\phi_y||\bs - \bs^\ast||)$ and $\rho(||\bs - \bs^\ast||;\phi_z) = \exp(-\phi_z||\bs - \bs^\ast||)$. Effective range estimates ($er_y$ and $er_z$) are again presented in the results tables to facilitate interpretation. The matrix $\bK = \bA\bA^\prime$ models the cross-covariance between the spatial random effects $u_y(\bs)$ and $u_z(\bs)$. The 2 x 2 parameter matrix $\bA$ is the lower triangular square-root of $\bK$. The diagonal and off-diagonal elements of $\bK$ capture the spatial processes variances and covariances, respectively. We assume the errors associated with the lidar sub-model are independent and normally distributed, i.e., $\bepsilon_z \overset{iid}{\sim} \mathcal{N}(\bzero,\tau_z^2\bI)$, where $\bepsilon_z = (\epsilon_z(\bs_1), \epsilon_z(\bs_2),..., \epsilon_z(\bs_m))^\prime$ and $\tau_z^2$ is a variance parameter to be estimated.

\subsubsection{Coregionalization$+$Tree Cover}
The \emph{Coregionalization} model described in model (\ref{eq:coreg}) can be extended to include wall-to-wall Landsat-derived tree cover information as such, 
\begin{equation}
\begin{bmatrix}y(\bs)\\z(\bs)\end{bmatrix} = \begin{bmatrix}\beta_{0y} + \beta_{1y}x(\bs)\\\beta_{0z} + \beta_{1z}x(\bs)\end{bmatrix} +\bA\begin{bmatrix}u_y(\bs)\\u_z(\bs)\end{bmatrix} + \begin{bmatrix}\epsilon_y(\bs)\\\epsilon_z(\bs)\end{bmatrix},
\end{equation}
where all terms with the exception of $\beta_{1z}$ have been previously defined. The lidar sub-model intercept, $\beta_{0y}$, and regression slope parameter, $\beta_{1y}$, together describe the linear relationship between the Landsat-based tree cover product, $x(\bs)$, and $z(\bs)$.

\subsection{Candidate model parameter estimation}
For all six candidate models, a Bayesian paradigm of statistical inference was pursued, which required us to specify prior distributions for all model parameters. Then inference proceeded by sampling from the posterior distribution of the parameters. We set prior distributions to be as vague as possible to minimize their influence on posterior inference. For the regression intercept and slope parameters, i.e., $\beta_{0y}$, $\beta_{1y}$, $\beta_{0z}$ and $\beta_{1z}$, we assumed a $\mathcal{N}(0,10\,000)$. Any spatial and non-spatial variance components, i.e., $\sigma^2_y$, $\tau^2_y$ and $\tau^2_z$, were assigned an inverse Gamma prior $IG(a,b)$. The $a$ hyperparameter was set equal to 2, which results in a prior distribution mean equal to $b$ and infinite variance. The $b$ hyperpriors were determined using preliminary semi-variograms fit to the residuals of the non-spatial models (\ref{eq:null}) and (\ref{eq:treecover}). The spatial decay parameters, $\phi_y$ and $\phi_z$ were assigned uniform priors with support over the geographic range of the study areas. The matrix $\bK$ was assigned an inverse Wishart prior with hyperparameter degrees of freedom set to 2 and diagonal scale matrix equaling $0.1\bI_2$, where $\bI_2$ is a 2 x 2 identity matrix. Algorithms for efficient estimation of parameters for all six candidate models are detailed in \citet{banerjee2014} and \citet{finley2014b}. All six models were fitted using the \texttt{spBayes} package written for the \texttt{R} statistical computing environment \citep{spBayes}. 

\subsection{Grid cell-level and areal estimation of aboveground biomass}\label{ppd}
After collecting a sufficient number of samples from the posterior distribution of a model's parameters via MCMC (following typical sampling and convergence diagnostics in \citet{gelman2013}), composition sampling was used to obtain samples from the posterior predictive distribution (PPD) of square-root AGB density at all grid cell locations within the study area \citep{banerjee2014}. Then each sample from the PPD was back-transformed (squared) to approximate spatially explicit AGB density PPDs ($\tilde\by = (\tilde{y}(\bs_1), \tilde{y}(\bs_2), \ldots, \tilde{y}(\bs_{n_0}))^{\prime}$, where $n_0$ is the number of prediction units, e.g., grid cells). Useful summaries of $\tilde{y}(\bs)$ for each grid cell, such as, median, standard deviation, or credible interval width, can be mapped to examine the spatial distribution of AGB density and associated uncertainty. $\tilde{\by}$ can also be summarized to estimate average AGB densities for arbitrary areal units. For example, to estimate mean AGB density for the full spatial domain we need to integrate the PPD of AGB density over the entire study region, i.e., $\tilde{y}(\mathcal{D}) = |\mathcal{D}|^{-1}\int_\mathcal{D}\tilde{y}(\bs)\text{d}\bs$ where $|\mathcal{D}|$ is the area of the domain $\mathcal{D}$. We can approximate this integral via MCMC integration using a fine grain systematic grid of $\tilde{y}(\bs)$ samples, i.e., $\tilde{y}(\mathcal{D}) \approx |\mathcal{D}|^{-1}\sum_{l = 1}^\mathcal{L}\tilde{y}(\bs_l)$, where $\mathcal{L}$ is the number of equally-spaced prediction grid points covering domain $\mathcal{D}$. To convert $\tilde{y}(\mathcal{D})$ to a PPD for total AGB, we multiply each sample by $|\mathcal{D}|$. This process also can be used to generate mean AGB density PPDs for sub-domains by integrating grid cell-level PPDs over redefined areal blocks, i.e., $|\mathcal{B}|^{-1}\int_\mathcal{B}\tilde{y}(\bs)\text{d}\bs$, where $\mathcal{B}$ is a sub-region of $\mathcal{D}$. See \citet[Chap. 7]{banerjee2014} for a more thorough discussion on summarizing PPDs over areal units. Total (mean) AGB estimates presented in the results tables were obtained by calculating the median of the total (mean) AGB PPD for the study area (labeled \emph{Est}). The standard deviation of the total (mean) AGB PPD is presented in the results tables as well and serves as a model-based uncertainty estimate for total (mean) AGB (labeled \emph{SD}). Relative standard deviations (labeled \emph{RSD} in the results tables) were calculated as $SD/Est*100\%$.

\subsection{Classical cokriging prediction}
Cokriging interpolation is commonly used in situations where multiple response variables need to be predicted across space and those responses are posited to co-varying across the domain. A cokriging approach is able to spatially predict the responses of interest accounting for spatial autocorrelation. Additionally, cokriging variance estimates can be calculated at prediction locations. Further, block-cokriging can generate small- and large-area estimates of the mean of a response. Given the similar capabilities of cokriging to the \emph{coregionalization} framework examined here and the documented use of cokriging in forestry and remote sensing, it is useful to conduct a comparison of the two approaches to potentially identify the merits and shortcomings of each \citep{meng2009, mutanga2006, tsui2013, wang2009}. Our interest is in understanding if either method produces more accurate point predictions and whether uncertainty intervals produced by each approach exhibit intended coverage. 

Cokriging interpolation of the BCEF, CPCRW, TNWR and TVSF data sets was implemented using the \texttt{gstat} package for the \texttt{R} statistical computing environment, similarly to the cokriging estimation approach detailed in \cite{tsui2013}. Specifically, the \texttt{variogram} function was used to develop semi- and cross-variograms for the square-root AGB and 90$^{\text{th}}$ percentile height responses. Subsequent exponential semi- and cross-variogram models were fitted using the \texttt{gstat} package's \texttt{fit.lmc} function. The \texttt{fit.lmc} function forces restrictions on semi- and cross-variogram parameter estimates to ensure resulting cokriging variance-covariance matrices are positive definite. Positive definiteness is a requirement for resulting cokriging prediction variances to be valid, i.e., variances be non-negative. One particularly strong restriction imposed by \texttt{fit.lmc} is that ranges for all semi- and cross-variogram models need to be equal. This is not a restriction for cokriging in general but approaches for building valid cross-variograms relaxing these types of restrictions can be complex and are not typically used in applications of cokriging in practice \citep{verhoef1998}. Once the models were fitted, the \texttt{gstat} package's \texttt{predict.krige} function was used to garner predictions and associated prediction variances for unobserved locations. 

\subsection{Cross-validation}
To examine grid cell-level predictive performance of the six candidate models, for each study site, a 10-fold holdout design was constructed by randomly assigning AGB plot observations to 10 approximately equal size groups. Square-root transformed AGB for each holdout group was sequentially predicted given model parameters estimated using data in the remaining nine groups. 10-fold holdout cross-validation root mean squared prediction error (\emph{CV-RMSE}) was calculated using back-transformed holdout posterior predicted medians and observed AGB for each model at all four sites. The model with the lowest \emph{CV-RMSE} was considered the \emph{best} grid cell-level predictor. The holdout predictions used for \emph{CV-RMSE} calculation were also used to generate the holdout residual semi-variograms presented in Fig.~\ref{varios}.

A similar 10-fold cross-validation procedure was used to obtain out-of-sample predictions using cokriging for the four sites. To avoid potential bias introduced due to back-transformation of cokriging predictions and variances, we elected to compare prediction accuracy and uncertainty interval coverage with the \emph{Coregionalization} model on the transformed ($\sqrt{Mg/ha}$) scale. Back-transformation in a frequentist setting is much more complicated than with the Bayesian models \citep{stow2006}. Empirical 95\% coverage probability was assessed by dividing the number of 95\% prediction uncertainty intervals containing $y(\bs)$ by the total number of observations at the site. For intervals exhibiting proper coverage, the empirical 95\% coverage probability should be near 0.95.

\section{Results and Discussion}

\subsection{Comparing non-spatial and spatial models}
Tables \ref{tab:bcef}, \ref{tab:cpcrw}, \ref{tab:tnwr} and \ref{tab:tvsf} present parameter posterior distribution summaries and prediction accuracy estimates for the six candidate models applied to the BCEF, CPCRW, TNWR and TVSF datasets. At all four sites, results show the \emph{Null} models to be more precise estimators of total (mean) AGB (labeled \emph{Est} in Tables \ref{tab:bcef}, \ref{tab:cpcrw}, \ref{tab:tnwr} and \ref{tab:tvsf}) than their \emph{Spatial} counterpart models, evidenced by lower total (mean) AGB standard deviations (labeled \emph{SD} in Tables \ref{tab:bcef}, \ref{tab:cpcrw}, \ref{tab:tnwr} and \ref{tab:tvsf}). However, the \emph{Null} models also show dramatically higher \emph{CV-RMSE} accuracy assessments compared to the \emph{Spatial} models at all sites, suggesting that the models including spatial random effects produced more accurate predictions.

\begin{figure}[!h]
  \includegraphics[width = .99\textwidth]{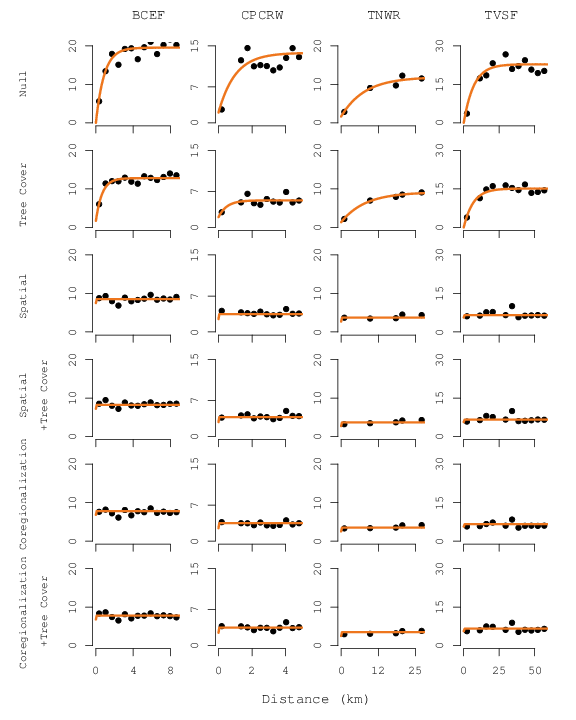}
  \caption{Holdout residual semi-variograms for the six candidate models at all four study sites.}\label{varios}
\end{figure}

The increased accuracy coupled with apparent decreased precision of the \emph{Spatial} model predictions highlight the role of distributional assumptions concerning the \emph{SD} and \emph{CV-RMSE} estimates. The \emph{CV-RMSE} metric employed here does not rely on distributional assumptions concerning stochastic model components because it only assesses the ability of the PPD median \emph{point estimate} to approximate observed AGB at plot locations---see \citet{efron1983} for discussion on robustness of cross-validation strategies to modeling assumptions. On the other hand, model-based summaries of PPDs, such as, the \emph{SD} precision metric presented in Tables \ref{tab:bcef}, \ref{tab:cpcrw}, \ref{tab:tnwr} and \ref{tab:tvsf}, heavily rely on modeling assumptions. Strong violations of distributional assumptions made during model-fitting can lead to unrealistic uncertainty estimates. Looking at the \emph{Null} model holdout residual semi-variograms in Fig.~\ref{varios}, we see that, at all four sites, there is strong residual spatial autocorrelation. This means that the \emph{Null} models likely violate the \emph{iid} error assumption imposed to conduct model-based inference about any parameters or predictions using standard deviations or other posterior distribution summaries. The \emph{SD} metrics for total (mean) AGB resulting from the \emph{Null} models are underestimated because the errors are falsely assumed to be independent of one another \citep{griffith2005}. The \emph{Spatial} model semi-variograms, however, show no signs of residual spatial structure, indicating that $w(\bs)$ has effectively absorbed any extraneous spatial variability (Fig.~\ref{varios}). Relaxing the \emph{iid} error assumption by incorporating spatial random effects allows for modelers to better interpret model-based uncertainty measures. Total (mean) AGB \emph{SD} estimates garnered via the \emph{Spatial} models are more reliable than the \emph{Null} model \emph{SD}s.

A similar phenomenon is observed when comparing the \emph{Tree Cover} and \emph{Spatial$+$Tree Cover} model predictive performance measures, although differences were smaller. Comparing the \emph{Null} and \emph{Tree Cover} holdout residual semi-variograms in Fig.~\ref{varios} shows that introducing the Landsat-derived tree cover variable absorbs a portion of the spatially structured variability in AGB at all four sites. However, spatial structure still exists in the holdout residuals of the \emph{Tree Cover} models. We argue that, due to positive residual spatial autocorrelation, \emph{SD} estimates are underestimated for the \emph{Tree Cover} models as well. Comparisons between the univariate non-spatial (\emph{Null} and \emph{Tree Cover}) and spatial (\emph{Spatial} and \emph{Spatial$+$Tree Cover}) models underscore the need to adhere to posited distributional assumptions concerning stochastic model components when attempting to interpret model-dependent measures of predictive performance, including grid cell-level and areally integrated PPDs. 

\begin{table}[p]
\caption{Bonanza Creek Experimental Forest (BCEF) candidate model parameter estimates, prediction accuracy metrics and total (mean) aboveground biomass (AGB) estimates. \emph{Est} = estimated total (mean) AGB with associated 95\% credible interval in parentheses. \emph{SD} = posterior predictive standard deviation for AGB total (mean) estimate. \emph{RSD} = relative standard deviation for total (mean) AGB estimate ($SD/Est*100\%$). \emph{CV-RMSE} = 10-fold holdout cross-validation root mean squared prediction error accuracy assessment.}\label{tab:bcef}
\centering
\resizebox{\textwidth}{!}{ 
{\renewcommand{\arraystretch}{1.5}
   \begin{tabular}{@{}clglglgl}\hline
  &                              & Null                  & Tree Cover           & Spatial             & Spatial$+$Tree Cover & Coregionalization   & Coregionalization$+$Tree Cover \\\hline
     \multirow{13}{*}{
       \rotatebox[origin = c]{90}{
         \specialcell{\small{Parameter posterior quantile summaries}\\[-.3em] \small{50\%} \small{(2.5\%, 97.5\%)}}
       }
     }
  &     $\beta_{0y}$              & 7.07 (6.56, 7.60)    & -0.41 (-1.93, 1.10)  & 7.05 (6.06, 7.81)    & 1.20 (-0.75, 3.19)  & 6.65 (5.90, 7.31)    & 0.97 (-0.56, 2.61)      \\
  &     $\beta_{1y}$              & ---                  & 0.10 (0.08, 0.12)    & ---                  & 0.08 (0.05, 0.10)   & ---                  & 0.08 (0.06, 0.10)       \\
  &     $\beta_{0z}$              & ---                  & ---                  & ---                  & ---                 & 9.35 (7.81, 11.13)   & 3.16 (1.81, 4.78)       \\
  &     $\beta_{1z}$              & ---                  & ---                  & ---                  & ---                 & ---                  & 0.09 (0.08, 0.11)       \\
  &     $\tau^2_y$                & 18.79 (15.90, 22.47) & 13.56 (11.50, 16.18) & 1.01 (0.27, 6.19)    & 3.43 (1.01, 6.96)   & 2.36 (1.58, 3.24)    & 1.81 (0.88, 3.08)       \\
  &     $\tau^2_z$                & ---                  & ---                  & ---                  & ---                 & 2.14 (1.54, 2.89)    & 2.03 (1.36, 2.82)       \\
  &     $\sigma^2_y$              & ---                  & ---                  & 17.83 (12.42, 23.46) & 10.19 (6.34, 14.83) & ---                  & ---                     \\
  &     $K[1,1]$                  & ---                  & ---                  & ---                  & ---                 & 20.71 (16.50, 25.57) & 13.60 (10.40, 17.98)    \\
  &     $K[1,2]$                  & ---                  & ---                  & ---                  & ---                 & 25.16 (21.25, 29.50) & 14.77 (11.56, 18.66)    \\
  &     $K[2,2]$                  & ---                  & ---                  & ---                  & ---                 & 41.39 (36.12, 48.35) & 29.51 (25.76, 34.92)    \\
  &     $\text{cor}(\bu_y,\bu_z)$ & ---                  & ---                  & ---                  & ---                 & 0.86 (0.77, 0.92)    & 0.74 (0.61, 0.86)       \\
  &     $er_y$ (km)               & ---                  & ---                  & 0.40 (0.23, 2.88)    & 0.32 (0.17, 3.03)   & 0.78 (0.59, 1.01)    & 0.36 (0.23, 0.54)       \\
  &     $er_z$ (km)               & ---                  & ---                  & ---                  & ---                 & 2.37 (1.37, 5.83)    & 2.15 (1.49, 3.61) \\\hline
%         \multirow{4}{*}{
%           \rotatebox[origin = c]{90}{
%             \specialcell{\small{Prediction Acc.}\\[-.3em]  \small{and Total Est.}}
%           }
%         }
  \multirow{3}{*}{\specialcell{Total\\[-.5em]AGB (Tg)}}  &$Est$& 1.328 (1.183, 1.495) & 1.307 (1.192, 1.446) & 1.323 (1.110, 1.582) & 1.286 (1.116, 1.485) & 1.323 (1.191, 1.484) & 1.310 (1.163, 1.469) \\
                                                         &$SD$ & 0.0809               & 0.0665               & 0.1235               & 0.0938               & 0.0787               & 0.0784               \\
                                                         &$RSD$& 6.09\%               & 5.09\%               & 9.33\%               & 7.29\%               & 5.95\%               & 5.98\%         \\\hline
  \multirow{3}{*}{\specialcell{Mean AGB\\[-.5em](Mg/ha)}}&$Est$& 68.92 (61.39, 77.59) & 67.86 (61.88, 75.08) & 68.67 (57.64, 82.12) & 66.77 (57.94, 77.11) & 68.66 (61.82, 77.04) & 67.99 (60.36, 76.25) \\
                                                         &$SD$ & 4.20                 & 3.45                 & 6.41                 & 4.87                 & 4.08                 & 4.07                 \\ 
                                                         &$RSD$& 6.09\%               & 5.09\%               & 9.33\%               & 7.29\%               & 5.95\%               & 5.98\%         \\\hline
  \multicolumn{2}{l}{\emph{CV-RMSE} (Mg/ha)}                   & 68.58                & 60.53                & 47.62                & 47.44                & 35.75                & 36.40          \\\hline
    \end{tabular}}}
\end{table}

\begin{table}[p]
\caption{Caribou-Poker Creeks Research Watershed (CPCRW) candidate model parameter estimates, prediction accuracy metrics and total (mean) aboveground biomass (AGB) estimates. \emph{Est} = estimated total (mean) AGB with associated 95\% credible interval in parentheses. \emph{SD} = posterior predictive standard deviation for AGB total (mean) estimate. \emph{RSD} = relative standard deviation for total (mean) AGB estimate ($SD/Est*100\%$). \emph{CV-RMSE} = 10-fold holdout cross-validation root mean squared prediction error accuracy assessment.}\label{tab:cpcrw}
\centering
\resizebox{\textwidth}{!}{ 
{\renewcommand{\arraystretch}{1.5}
   \begin{tabular}{@{}clglglgl}\hline
  &                               & Null                & Tree Cover          & Spatial            & Spatial$+$Tree Cover & Coregionalization    & Coregionalization$+$Tree Cover  \\\hline
     \multirow{13}{*}{
       \rotatebox[origin = c]{90}{
         \specialcell{\small{Parameter posterior quantile summaries}\\[-.3em] \small{50\%} \small{(2.5\%, 97.5\%)}}
       }
     }
  &     $\beta_{0y}$              & 5.68 (5.13, 6.24)   & -0.11 (-1.04, 0.81) & 5.65 (4.72, 6.75)  & 1.00 (-0.35, 2.64)   & 4.97 (3.92, 5.88)    & 2.06 (0.76, 3.54)       \\
  &     $\beta_{1y}$              & ---                 & 0.09 (0.08, 0.11)   & ---                & 0.08 (0.05, 0.09)    & ---                  & 0.05 (0.03, 0.07)       \\
  &     $\beta_{0z}$              & ---                 & ---                 & ---                & ---                  & 5.25 (3.14, 7.15)    & 0.66 (-1.28, 2.24)      \\
  &     $\beta_{1z}$              & ---                 & ---                 & ---                & ---                  & ---                  & 0.08 (0.07, 0.10)       \\
  &     $\tau^2_y$                & 11.42 (9.16, 14.43) & 5.16 (4.13, 6.55)   & 1.65 (0.76, 2.62)  & 2.35 (0.80, 3.76)    & 1.66 (1.22, 2.21)    & 1.99 (1.25, 2.75)       \\
  &     $\tau^2_z$                & ---                 & ---                 & ---                & ---                  & 1.82 (1.28, 2.52)    & 2.15 (1.51, 2.91)       \\
  &     $\sigma^2_y$              & ---                 & ---                 & 8.96 (6.10, 13.58) & 2.98 (1.20, 5.41)    & ---                  & ---                     \\
  &     $K[1,1]$                  & ---                 & ---                 & ---                & ---                  & 9.63 (6.83, 12.95)   & 4.86 (2.59, 8.59)       \\
  &     $K[1,2]$                  & ---                 & ---                 & ---                & ---                  & 14.03 (10.56, 18.38) & 7.10 (4.74, 10.03)      \\
  &     $K[2,2]$                  & ---                 & ---                 & ---                & ---                  & 26.32 (20.46, 34.39) & 14.60 (11.61, 19.55)    \\
  &     $\text{cor}(\bu_y,\bu_z)$ & ---                 & ---                 & ---                & ---                  & 0.89 (0.81, 0.94)    & 0.85 (0.71, 0.92)       \\
  &     $er_y$ (km)               & ---                 & ---                 & 1.28 (0.55, 2.91)  & 0.33 (0.10, 2.19)    & 1.96 (1.40, 2.95)    & 1.20 (0.76, 1.92)       \\
  &     $er_z$ (km)               & ---                 & ---                 & ---                & ---                  & 2.49 (1.66, 6.35)    & 2.31 (1.32, 6.01) \\\hline
%         \multirow{4}{*}{
%           \rotatebox[origin = c]{90}{
%             \specialcell{\small{Prediction Acc.}\\[-.3em]  \small{and Total Est.}}
%           }
%         }
  \multirow{3}{*}{\specialcell{Total\\[-.5em]AGB (Tg)}}  &$Est$& 0.464 (0.396, 0.530) & 0.462 (0.414, 0.514) & 0.457 (0.380, 0.554) & 0.444 (0.387, 0.506) & 0.413 (0.380, 0.449) & 0.403 (0.362, 0.442) \\
                                                         &$SD$ & 0.0355               & 0.0250               & 0.0444               & 0.0316               & 0.0173               & 0.0199               \\
                                                         &$RSD$& 7.65\%               & 5.41\%               & 9.73\%               & 7.12\%               & 4.18\%               & 4.94\%         \\\hline
  \multirow{3}{*}{\specialcell{Mean AGB\\[-.5em](Mg/ha)}}&$Est$& 43.87 (37.45, 50.17) & 43.66 (39.20, 48.60) & 43.27 (35.93, 52.39) & 41.99 (36.64, 47.89) & 39.07 (35.98, 42.43) & 38.16 (34.27, 41.83) \\
                                                         &$SD$ & 3.35                 & 2.36                 & 4.20                 & 2.99                 & 1.63                 & 1.88                 \\
                                                         &$RSD$& 7.65\%               & 5.41\%               & 9.73\%               & 7.12\%               & 4.18\%               & 4.94\%         \\\hline
  \multicolumn{2}{l}{\emph{CV-RMSE} (Mg/ha)}                   & 41.55                & 27.41                & 23.88                & 24.06                & 20.78                & 21.19          \\\hline
    \end{tabular}}}
\end{table}

\begin{table}[p]
\caption{Tetlin National Wildlife Refuge (TNWR) candidate model parameter estimates, prediction accuracy metrics and total (mean) aboveground biomass (AGB) estimates. \emph{Est} = estimated total (mean) AGB with associated 95\% credible interval in parentheses. \emph{SD} = posterior predictive standard deviation for AGB total (mean) estimate. \emph{RSD} = relative standard deviation for total (mean) AGB estimate ($SD/Est*100\%$). \emph{CV-RMSE} = 10-fold holdout cross-validation root mean squared prediction error accuracy assessment.}\label{tab:tnwr}
\centering
\resizebox{\textwidth}{!}{ 
{\renewcommand{\arraystretch}{1.5}
   \begin{tabular}{@{}clglglgl}\hline
  &                               & Null                 & Tree Cover          & Spatial             & Spatial$+$Tree Cover & Coregionalization   & Coregionalization$+$Tree Cover  \\\hline
     \multirow{13}{*}{
       \rotatebox[origin = c]{90}{
         \specialcell{\small{Parameter posterior quantile summaries}\\[-.3em] \small{50\%} \small{(2.5\%, 97.5\%)}}
       }
     }
  &     $\beta_{0y}$              & 4.19 (3.46, 4.92)    & 0.71 (-0.47, 1.92)  & 4.23 (3.15, 5.40)   & 1.36 (-0.04, 3.04)   & 3.87 (3.13, 4.63)    & 0.98 (-0.05, 1.91)       \\
  &     $\beta_{1y}$              & ---                  & 0.07 (0.05, 0.10)   & ---                 & 0.06 (0.04, 0.09)    & ---                  & 0.07 (0.05, 0.09)        \\
  &     $\beta_{0z}$              & ---                  & ---                 & ---                 & ---                  & 3.09 (2.38, 3.68)    & -0.39 (-1.00, 0.14)      \\
  &     $\beta_{1z}$              & ---                  & ---                 & ---                 & ---                  & ---                  & 0.08 (0.07, 0.09)        \\
  &     $\tau^2_y$                & 14.73 (11.36, 19.61) & 10.43 (8.06, 13.80) & 0.60 (0.18, 2.28)   & 1.32 (0.51, 3.80)    & 0.79 (0.20, 2.02)    & 0.74 (0.20, 1.62)        \\
  &     $\tau^2_z$                & ---                  & ---                 & ---                 & ---                  & 1.96 (1.39, 2.70)    & 2.24 (1.63, 2.95)        \\
  &     $\sigma^2_y$              & ---                  & ---                 & 12.90 (8.57, 21.39) & 8.59 (5.46, 12.98)   & ---                  & ---                      \\
  &     $K[1,1]$                  & ---                  & ---                 & ---                 & ---                  & 17.70 (12.69, 24.90) & 11.50 (8.12, 15.24)      \\
  &     $K[1,2]$                  & ---                  & ---                 & ---                 & ---                  & 13.58 (11.27, 16.92) & 8.48 (7.05, 10.16)       \\
  &     $K[2,2]$                  & ---                  & ---                 & ---                 & ---                  & 15.31 (13.57, 17.77) & 9.05 (7.80, 10.73)       \\
  &     $\text{cor}(\bu_y,\bu_z)$ & ---                  & ---                 & ---                 & ---                  & 0.83 (0.75, 0.89)    & 0.83 (0.76, 0.89)        \\
  &     $er_y$ (km)               & ---                  & ---                 & 0.36 (0.19, 0.72)   & 0.28 (0.15, 0.83)    & 0.57 (0.44, 0.74)    & 0.39 (0.30, 0.50)        \\
  &     $er_z$ (km)               & ---                  & ---                 & ---                 & ---                  & 5.23 (2.96, 9.71)    & 5.19 (2.80, 19.17) \\\hline
%         \multirow{4}{*}{
%           \rotatebox[origin = c]{90}{
%             \specialcell{\small{Prediction Acc.}\\[-.3em]  \small{and Total Est.}}
%           }
%         }
  \multirow{3}{*}{\specialcell{Total\\[-.5em]AGB (Tg)}}  &$Est$& 8.654 (7.091, 11.007) & 8.133 (6.716, 10.126) & 8.702 (6.189, 12.429) & 7.996 (5.688, 10.716) & 9.082 (7.662, 11.176) & 8.381 (7.183, 9.682)\\
                                                         &$SD$ & 1.0108                & 0.8610                & 1.6128                & 1.1950                & 0.9482                & 0.6477              \\
                                                         &$RSD$& 11.68\%               & 10.59\%               & 18.53\%               & 14.94\%               & 10.44\%               & 7.73\%        \\\hline
  \multirow{3}{*}{\specialcell{Mean AGB\\[-.5em](Mg/ha)}}&$Est$& 32.39 (26.54, 41.19)  & 30.44 (25.13, 37.90)  & 32.56 (23.16, 46.51)  & 29.92 (21.29, 40.10)  & 33.99 (28.67, 41.82)  & 31.36 (26.88, 36.23)\\
                                                         &$SD$ & 3.78                  & 3.22                  & 6.04                  & 4.47                  & 3.55                  & 2.42                \\
                                                         &$RSD$& 11.68\%               & 10.59\%               & 18.53\%               & 14.94\%               & 10.44\%               & 7.73\%        \\\hline
  \multicolumn{2}{l}{\emph{CV-RMSE} (Mg/ha)}                   & 51.89                 & 44.70                 & 37.37                 & 38.01                 & 34.03                 & 30.89         \\\hline
    \end{tabular}}}
\end{table}

\begin{table}[p]
\caption{Tanana Valley State Forest (TVSF) candidate model parameter estimates, prediction accuracy metrics and total (mean) aboveground biomass (AGB) estimates. \emph{Est} = estimated total (mean) AGB with associated 95\% credible interval in parentheses. \emph{SD} = posterior predictive standard deviation for AGB total (mean) estimate. \emph{RSD} = relative standard deviation for total (mean) AGB estimate ($SD/Est*100\%$). \emph{CV-RMSE} = 10-fold holdout cross-validation root mean squared prediction error accuracy assessment.}\label{tab:tvsf}
\centering
\resizebox{\textwidth}{!}{ 
{\renewcommand{\arraystretch}{1.5}
   \begin{tabular}{@{}clglglgl}\hline
  &                               & Null                    & Tree Cover         & Spatial               & Spatial$+$Tree Cover & Coregionalization    & Coregionalization$+$Tree Cover  \\\hline
     \multirow{13}{*}{
       \rotatebox[origin = c]{90}{
         \specialcell{\small{Parameter posterior quantile summaries}\\[-.3em] \small{50\%} \small{(2.5\%, 97.5\%)}}
       }
     }
  &     $\beta_{0y}$              & 7.10 (6.55, 7.63)      & 0.48 (-0.97, 1.92)   & 7.09 (6.19, 8.02)    & 2.14 (0.24, 4.42)    & 7.41 (6.74, 8.01)    & 2.09 (0.74, 3.52)       \\
  &     $\beta_{1y}$              & ---                    & 0.09 (0.07, 0.11)    & ---                  & 0.07 (0.04, 0.10)    & ---                  & 0.08 (0.06, 0.09)       \\
  &     $\beta_{0z}$              & ---                    & ---                  & ---                  & ---                  & 8.35 (7.68, 9.06)    & 3.03 (2.50, 3.68)       \\
  &     $\beta_{1z}$              & ---                    & ---                  & ---                  & ---                  & ---                  & 0.08 (0.07, 0.09)       \\
  &     $\tau^2_y$                & 19.53 (16.53, 23.31)   & 14.47 (12.19, 17.30) & 0.91 (0.21, 3.54)    & 1.47 (0.47, 3.51)    & 0.99 (0.35, 1.53)    & 0.68 (0.32, 1.85)       \\
  &     $\tau^2_z$                & ---                    & ---                  & ---                  & ---                  & 2.61 (2.03, 3.20)    & 2.48 (2.05, 2.96)       \\
  &     $\sigma^2_y$              & ---                    & ---                  & 18.60 (14.04, 24.08) & 13.15 (9.67, 18.23)  & ---                  & ---                     \\
  &     $K[1,1]$                  & ---                    & ---                  & ---                  & ---                  & 23.80 (18.93, 29.98) & 17.38 (13.50, 21.67)    \\
  &     $K[1,2]$                  & ---                    & ---                  & ---                  & ---                  & 22.61 (19.30, 26.37) & 16.48 (14.50, 18.90)    \\
  &     $K[2,2]$                  & ---                    & ---                  & ---                  & ---                  & 40.00 (36.73, 43.89) & 27.86 (25.89, 29.63)    \\
  &     $\text{cor}(\bu_y,\bu_z)$ & ---                    & ---                  & ---                  & ---                  & 0.73 (0.67, 0.80)    & 0.75 (0.70, 0.81)       \\
  &     $er_y$ (km)               & ---                    & ---                  & 0.56 (0.35, 1.28)    & 0.42 (0.26, 0.87)    & 0.73 (0.60, 0.91)    & 0.47 (0.40, 0.58)       \\
  &     $er_z$ (km)               & ---                    & ---                  & ---                  & ---                  & 7.10 (5.14, 11.17)   & 4.49 (3.39, 6.72) \\\hline
%         \multirow{4}{*}{
%           \rotatebox[origin = c]{90}{
%             \specialcell{\small{Prediction Acc.}\\[-.3em]  \small{and Total Est.}}
%           }
%         }
  \multirow{3}{*}{\specialcell{Total\\[-.5em]AGB (Tg)}}  &$Est$& 49.05 (43.47, 54.73) & 47.18 (42.70, 51.86) & 49.31 (40.31, 60.10) & 46.53 (39.71, 54.81) & 55.82 (50.22, 61.59) & 52.54 (49.31, 57.12) \\
                                                         &$SD$ & 2.8404               & 2.4088               & 5.0405               & 4.1593               & 2.9045               & 2.0866               \\
                                                         &$RSD$& 5.79\%               & 5.11\%               & 10.28\%              & 8.94\%               & 5.20\%               & 3.97\%         \\\hline
  \multirow{3}{*}{\specialcell{Mean AGB\\[-.5em](Mg/ha)}}&$Est$& 70.39 (62.38, 78.55) & 67.70 (61.28, 74.42) & 70.35 (57.85, 86.25) & 66.77 (56.99, 78.66) & 80.11 (72.07, 88.39) & 75.40 (70.76, 81.98) \\
                                                         &$SD$ & 4.08                 & 3.46                 & 7.23                 & 5.96                 & 4.17                 & 2.0866               \\
                                                         &$RSD$& 5.79\%               & 5.11\%               & 10.28\%              & 8.94\%               & 5.20\%               & 3.97\%         \\\hline
  \multicolumn{2}{l}{\emph{CV-RMSE} (Mg/ha)}                   & 72.37                & 63.46                & 42.68                & 42.15                & 31.00                & 28.95          \\\hline
    \end{tabular}}}
\end{table}

\subsection{Comparing spatial and coregionalization models}
At all four sites, the lidar-informed \emph{Coregionalization} models outperformed their counterpart lidar-free \emph{Spatial} models by producing lower \emph{CV-RMSE} accuracy assessments and total (mean) AGB \emph{SD} metrics. Similarly, the \emph{Coregionalization$+$Tree Cover} models produced better accuracy measures than the \emph{Spatial$+$Tree Cover} models. Prediction gains resulting from the incorporation of lidar information are not surprising considering that lidar-derived measures of forest height are strongly related to AGB. It is encouraging, however, that this relationship produces appreciable gains in total (mean) AGB estimation precision at such low lidar sampling intensities---see Section \ref{rsdata} for description of the lidar dataset subsampling. Tables \ref{tab:bcef}, \ref{tab:cpcrw}, \ref{tab:tnwr} and \ref{tab:tvsf} show the correlation between the spatial random effects for the AGB and lidar sub-models (labeled \emph{$\text{cor}(\bu_y,\bu_z)$} in Tables \ref{tab:bcef}, \ref{tab:cpcrw}, \ref{tab:tnwr} and \ref{tab:tvsf}). Median point estimates of correlation range between 0.73 and 0.89 indicating strong correspondence between square-root transformed AGB density and 90$^{\text{th}}$ percentile height. Since the relationship between the AGB and lidar responses is so substantial, considerable improvements in predictive performance and total (mean) AGB estimation precision are realized. At all sites, relative standard deviations (labeled \emph{RSD} in Tables \ref{tab:bcef}, \ref{tab:cpcrw}, \ref{tab:tnwr} and \ref{tab:tvsf}) for the \emph{Coregionalization} models are nearly half that of the \emph{Spatial} models. 

At the sites with sparser field and lidar sampling, i.e., TVSF and TNWR, we see further gains in grid cell-level prediction accuracy and total (mean) AGB certainty when the Landsat-based tree cover product is included---evidenced by \emph{Coregionalization$+$Tree Cover} models having lower \emph{RSD} and \emph{CV-RMSE} estimates than counterpart \emph{Coregionalization} frameworks. Examining effective range estimates (labeled $er_y$ and $er_z$ in Tables \ref{tab:bcef}, \ref{tab:cpcrw}, \ref{tab:tnwr} and \ref{tab:tvsf}) offers insight concerning this phenomenon. Effective range estimates help us understand how far-reaching the underlying spatial dependence structure is in the responses, i.e., $y(\bs)$ and $z(\bs)$, after accounting for covariates. The longer the effective range, the larger the proximate neighborhood each grid cell draws information from. The $er_z$ median point estimates for the TNWR and TVSF \emph{Coregionalization} and \emph{Coregionalization$+$Tree Cover} models extend less than the 10 km separation distance between lidar flight-lines at those sites (4.49 km $< er_z < 7.10$ km). However, the effective range estimates for the corresponding models exceed the average flight-line separation distances of $< 1.5$ km at BCEF and CPCRW (2.15 km $< er_z < 2.41$ km). On average,  grid cell-level predictions at BCEF and CPCRW are borrowing more proximate lidar data to inform prediction because there is more available within the effective range of spatial dependence at these sites. Since field plots are also closer together at CPCRW and BCEF compared to TVSF and TNWR, grid cell predictions are potentially tapping into more information provided by nearby AGB observations as well. Any AGB explanatory power offered by the Landsat tree cover product is already explained by simply borrowing strength from nearby field and lidar data to inform prediction at BCEF and CPCRW. However, since field and lidar data are much farther apart at TVSF and TNWR, we do see benefit from the wall-to-wall information provided by the Landsat-based tree cover product.

\subsection{Comparing Coregionalization and Coregionalization$+$Tree Cover mapped predictions}
\begin{figure}[!hb]
  \centering
  \subfloat[BCEF: Coregionalization predictions]{\includegraphics[width = .5\textwidth]{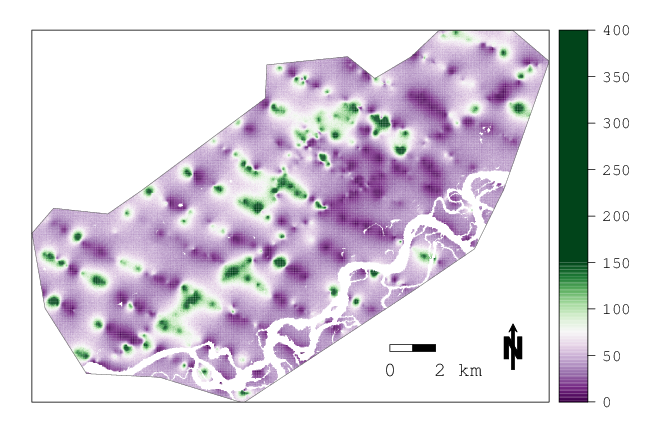}}
  \subfloat[BCEF: Coregionalization prediction standard deviations]{\includegraphics[width = .5\textwidth]{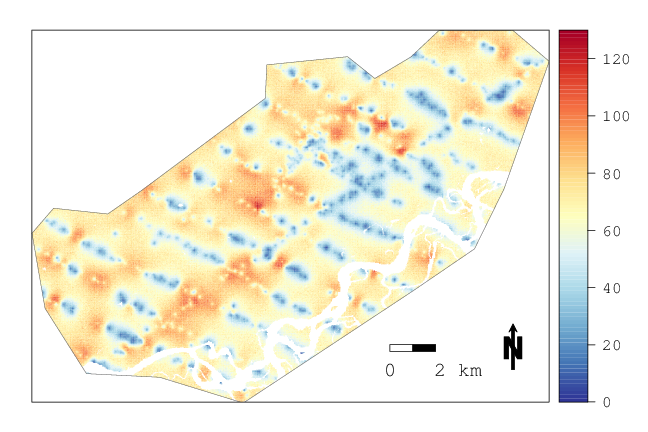}}\\
  \subfloat[CPCRW: Coregionalization predictions]{\includegraphics[width = .5\textwidth]{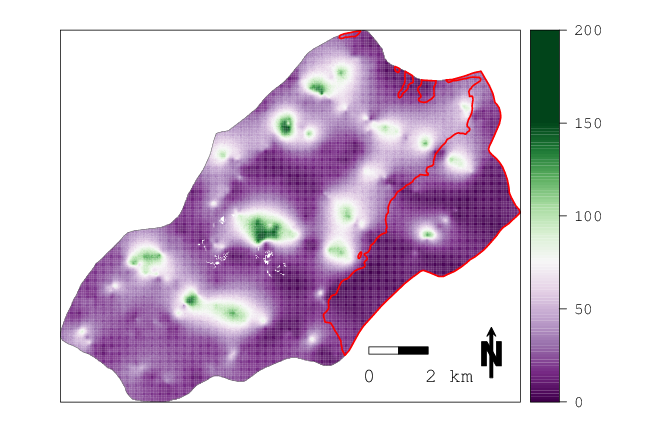}}
  \subfloat[CPCRW: Coregionalization prediction standard deviations]{\includegraphics[width = .5\textwidth]{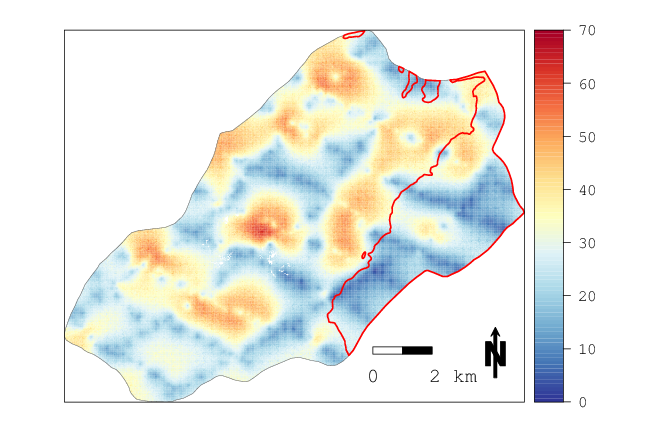}}
  \caption{Mapped grid cell-level predictions of aboveground biomass density (a and c) and associated standard deviations (b and d) (Mg/ha) using the \emph{Coregionalization} model for Bonanza Creek Experimental Forest (BCEF) and Caribou-Poker Creeks Research Watershed (CPCRW). The red polygon boundary shown in c and d defines the area burned during the 2004 Boundary Fire at CPCRW.}\label{pred5-maps1}
\end{figure}

\begin{figure}[!hb]
  \subfloat[TNWR: Coregionalization predictions]{\includegraphics[width = .5\textwidth]{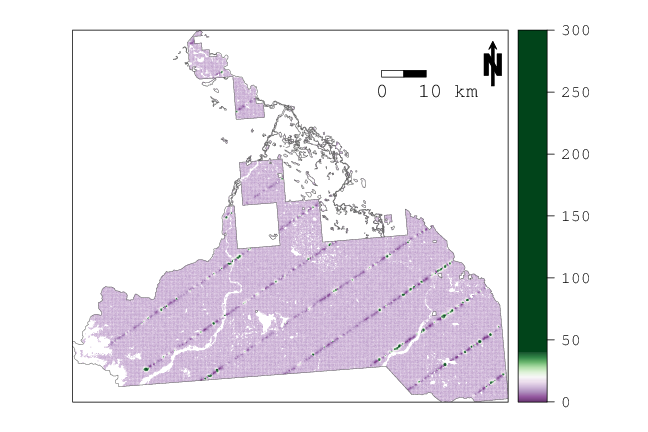}}
  \subfloat[TNWR: Coregionalization prediction standard deviations]{\includegraphics[width = .5\textwidth]{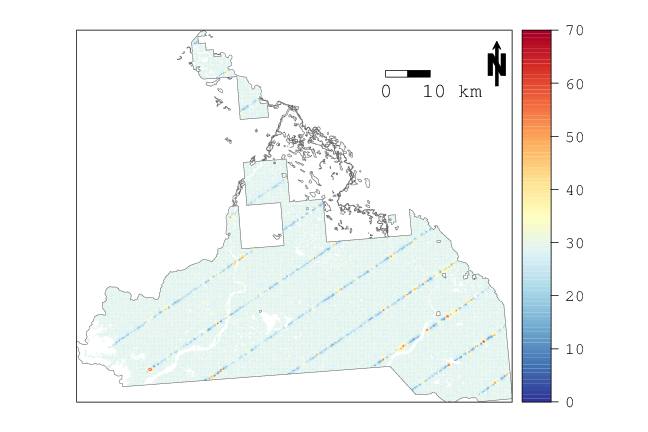}}\\
  \subfloat[TVSF: Coregionalization predictions]{\includegraphics[width = .5\textwidth]{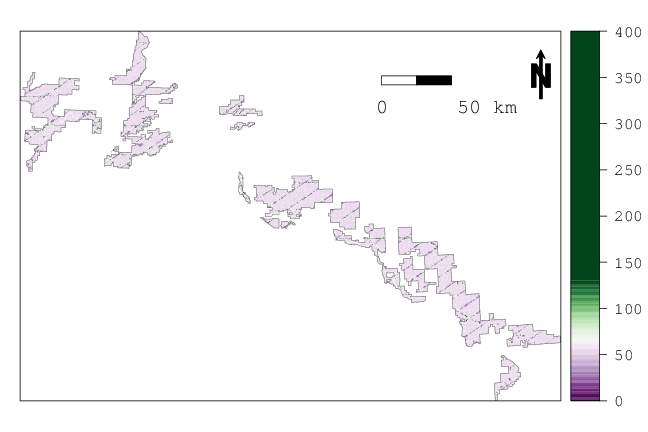}}
  \subfloat[TVSF: Coregionalization prediction standard deviations]{\includegraphics[width = .5\textwidth]{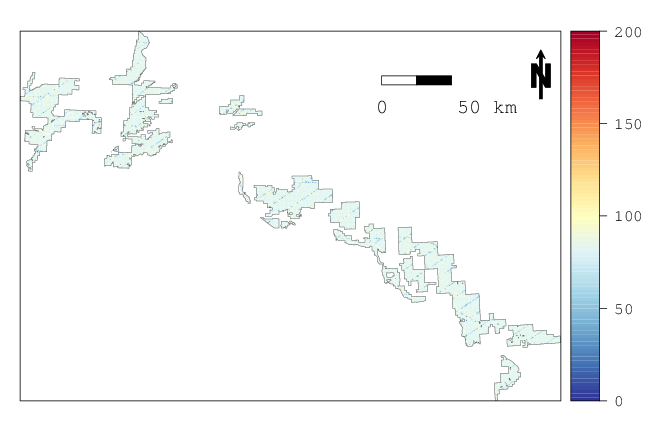}}
  \caption{Mapped grid cell-level predictions of aboveground biomass density (a and c) and associated standard deviations (b and d) (Mg/ha) using the \emph{Coregionalization} model for Tetlin National Wildlife Refuge (TNWR) and Tanana Valley State Forest (TVSF).}\label{pred5-maps2}
\end{figure}

\begin{figure}[!hb]
  \centering
  \subfloat[BCEF: Coregionalization$+$Tree Cover predictions]{\includegraphics[width = .5\textwidth]{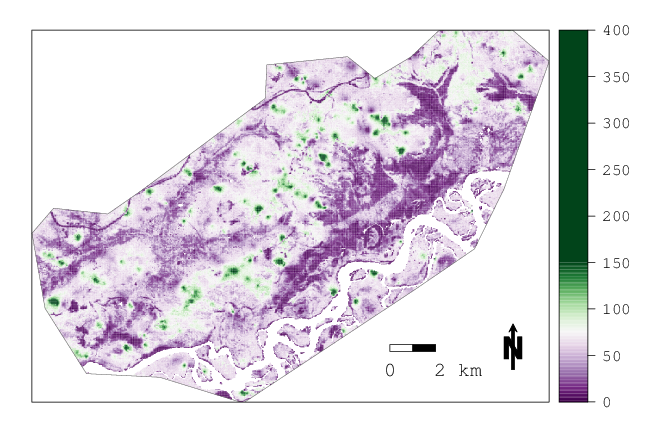}}
  \subfloat[BCEF: Coregionalization$+$Tree Cover prediction standard deviations]{\includegraphics[width = .5\textwidth]{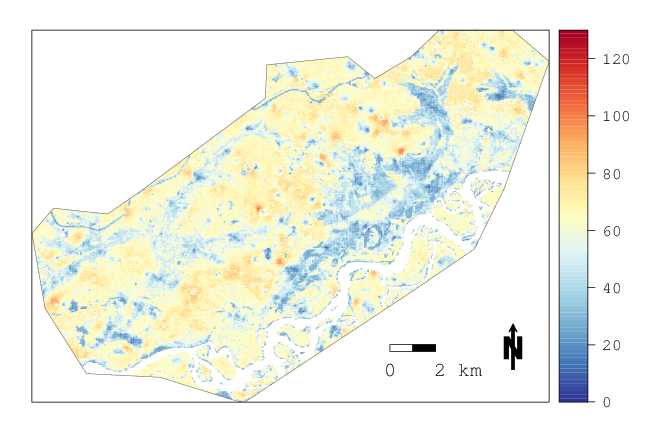}}\\
  \subfloat[CPCRW: Coregionalization$+$Tree Cover predictions]{\includegraphics[width = .5\textwidth]{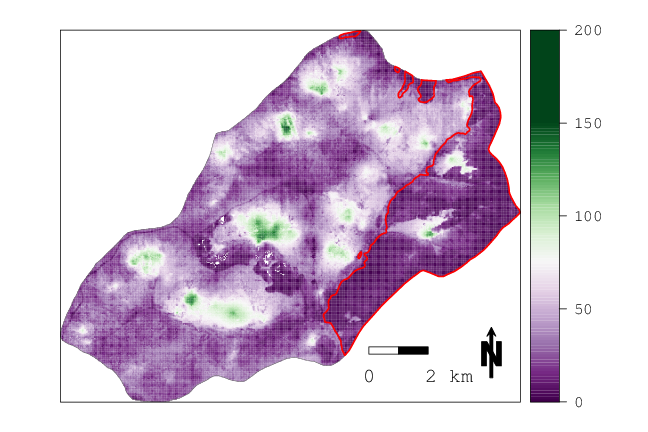}}
  \subfloat[CPCRW: Coregionalization$+$Tree Cover prediction standard deviations]{\includegraphics[width = .5\textwidth]{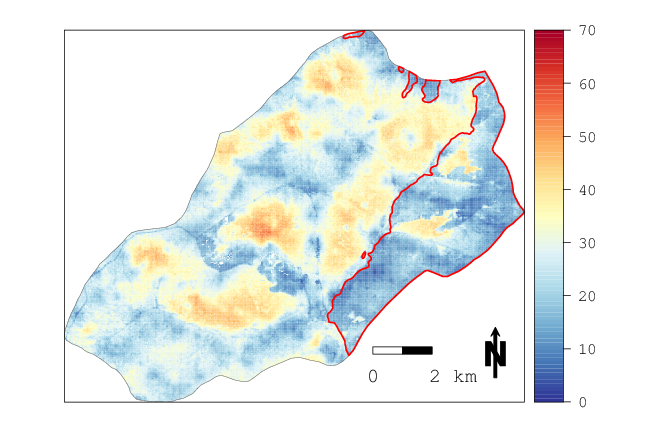}}
  \caption{Mapped grid cell-level predictions of aboveground biomass density (a and c) and associated standard deviations (b and d) (Mg/ha) using the \emph{Coregionalization$+$Tree Cover} model for Bonanza Creek Experimental Forest (BCEF) and Caribou-Poker Creeks Research Watershed (CPCRW). The red polygon boundary shown in c and d defines the area burned during the 2004 Boundary Fire at CPCRW.}\label{pred6-maps1}
\end{figure}

\begin{figure}[!hb]
  \subfloat[TNWR: Coregionalization$+$Tree Cover predictions]{\includegraphics[width = .5\textwidth]{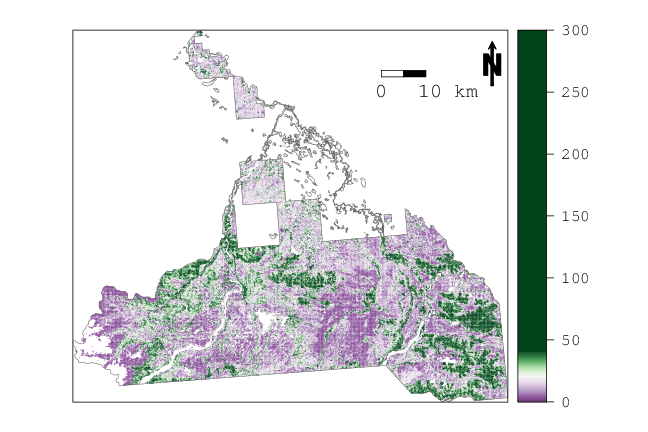}}
  \subfloat[TNWR: Coregionalization$+$Tree Cover prediction standard deviations]{\includegraphics[width = .5\textwidth]{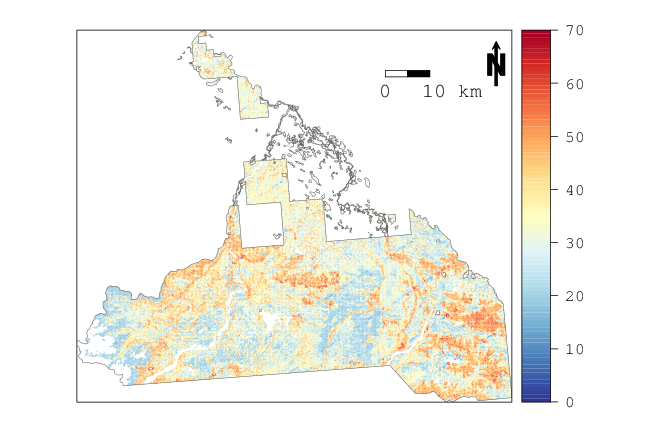}}\\
  \subfloat[TVSF: Coregionalization$+$Tree Cover predictions]{\includegraphics[width = .5\textwidth]{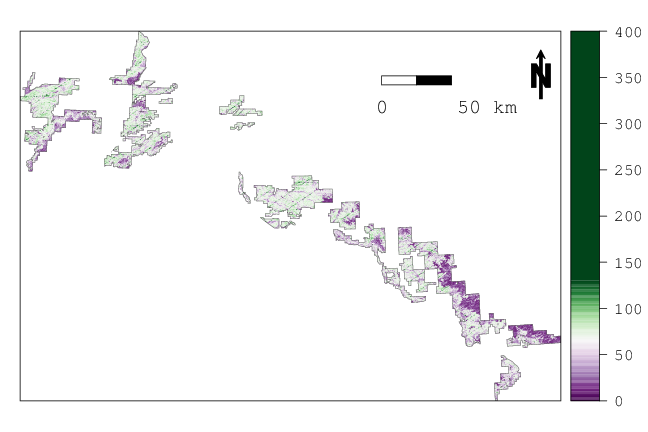}}
  \subfloat[TVSF: Coregionalization$+$Tree Cover prediction standard deviations]{\includegraphics[width = .5\textwidth]{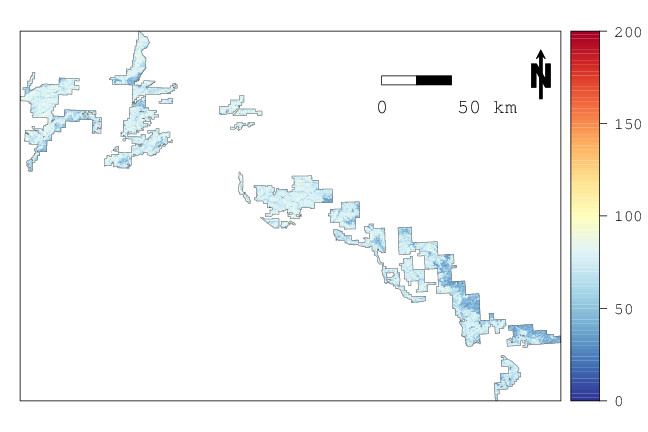}}
  \caption{Mapped grid cell-level predictions of aboveground biomass density (a and c) and associated standard deviations (b and d) (Mg/ha) using the \emph{Coregionalization$+$Tree Cover} model for Tetlin National Wildlife Refuge (TNWR) and Tanana Valley State Forest (TVSF).}\label{pred6-maps2}
\end{figure}

Figs. \ref{pred5-maps1}, \ref{pred5-maps2}, \ref{pred6-maps1} and \ref{pred6-maps2} show maps of PPD median point estimates and standard deviations for AGB density using the \emph{Coregionalization} and \emph{Coregionalization$+$Tree Cover} models. The flight-lines where lidar was collected are easily discernible in the \emph{Coregionalization} model prediction maps (Figs. \ref{pred5-maps1} and \ref{pred5-maps2}). We see differentiation of AGB within lidar strips and near field plot locations. As one moves away from regions where information was collected, predicted AGB densities retreat to the global mean. Additionally, grid cell-level uncertainty maps show a trend of higher predictive precision in close proximity to field and lidar observations compared to areas farther away. Even though the \emph{Coregionalization} models provide reliable estimates of total (mean) AGB and show favorable \emph{CV-RMSE} accuracy assessments, the resulting prediction maps leave something to be desired. Having no information outside the flight-lines, the coregionalization models only rely on borrowing strength from nearby observations to inform prediction between strips. The addition of the wall-to-wall Landsat tree cover product allows for better differentiation of median point predictions outside lidar strips, leading to maps that are potentially more useful for assessing the spatial distribution of AGB across the landscape. At TVSF and TNWR, \emph{CV-RMSE} estimates suggest that the addition of the Landsat tree cover product leads to better map accuracy. However, at BCEF and CPCRW, the inclusion of Landsat tree cover appears to only offer cosmetic adjustments to resulting maps, not actual increases in prediction accuracy. It seems that the intensity of the field and lidar sample at BCEF and CPCRW mixed with constructing models to borrow information from neighboring locations offered more predictive advantage than simply including Landsat tree cover as a covariate.

\subsection{Watershed unit AGB estimates at Caribou-Poker Creeks Research Watershed}
\begin{figure}[!hb]
  \subfloat{\includegraphics[width = .6\textwidth]{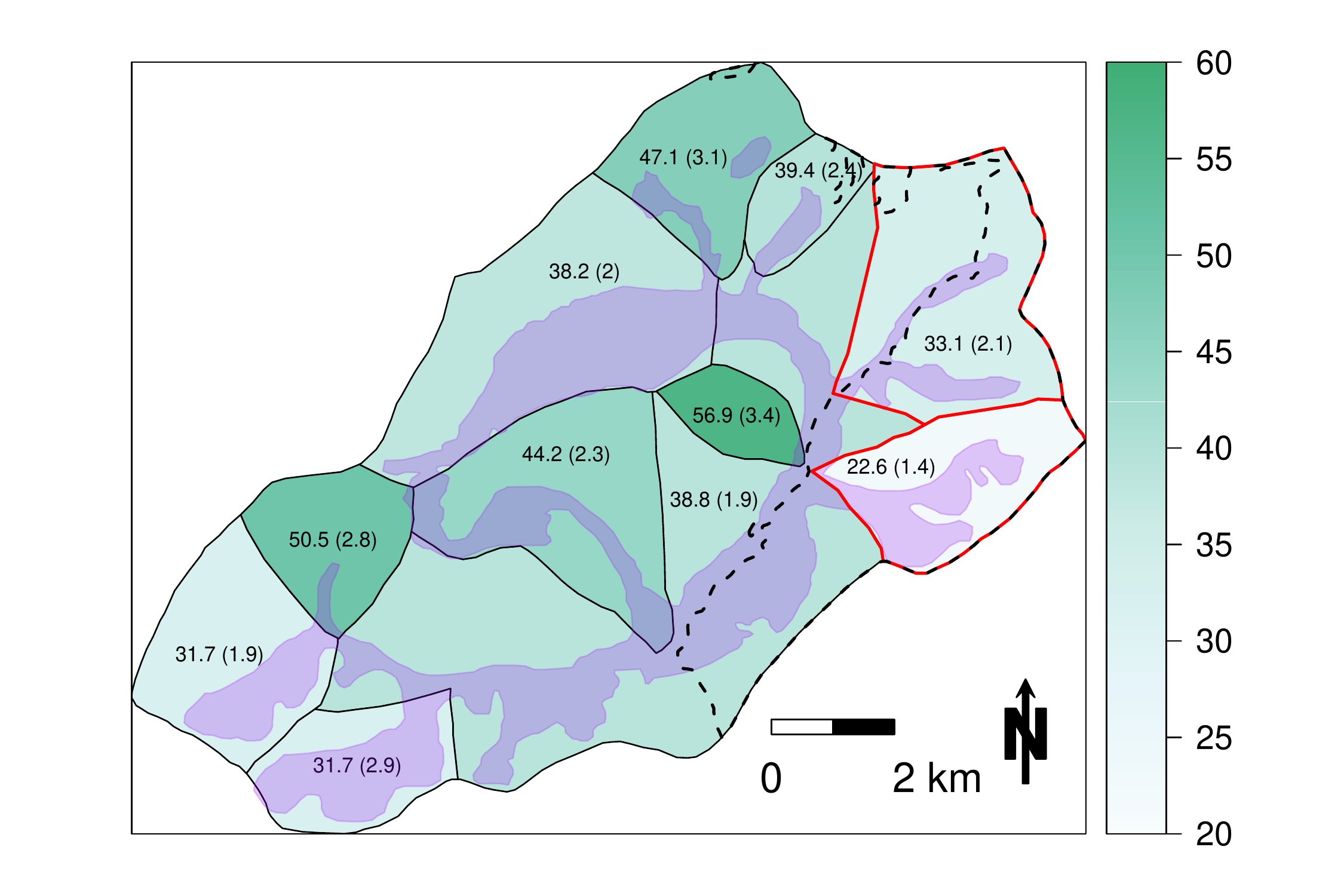}}
  \subfloat{\includegraphics[width = .4\textwidth]{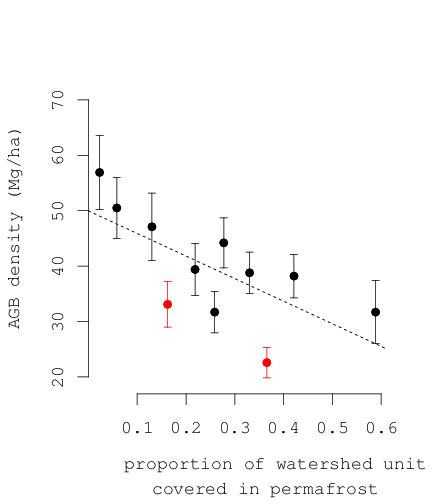}}
  \caption{Left figure maps watershed unit-level mean aboveground biomass (AGB) (Mg/ha) with associated standard deviations in parentheses using the \emph{Coregionalization$+$Tree Cover} model for Caribou-Poker Creeks Research Watershed. Solid polygon boundaries delineate watershed units. Translucent purple polygon identifies area covered in permafrost. The black dashed line delineates the border of the 2004 Boundary Fire. The red outlined watershed units are identified as the two units with the highest proportion of area burned during the 2004 Boundary Fire. The right figure shows a scatter plot highlighting the relationship between the proportion of permafrost in a watershed unit and AGB density. The dashed trend-line shows the general relationship between AGB density and permafrost proportion. The points shown in red correspond to the red outlined watershed units in the left figure.}\label{ws-map}
\end{figure}

Fig.~\ref{ws-map} shows estimated mean AGB densities for the 11 watershed units at CPCRW predicted using the \emph{Coregionalization$+$Tree Cover} model with a permafrost polygon layer overlaid. Average AGB density was negatively correlated with permafrost presence at the watershed unit level. By appropriately summarizing the model's PPD surface, it is possible to obtain an estimate of the correlation between AGB density and permafrost proportion at the watershed unit scale while accounting for uncertainty in the AGB density predictions. By calculating the correlation between the proportion of the watershed unit covered in permafrost with each MCMC sample of average AGB density, we can obtain a PPD summary characterizing their degree of correlation. The median correlation point estimate representing the relationship between permafrost and watershed-unit AGB density is $-0.685$. The 95 percent upper and lower credible bounds for this estimate are $-0.587$ and $-0.764$, respectively. Since this interval does not include zero, we can conclude that permafrost presence is negatively related to AGB density with 95 percent certainty. 

We can also explore potential causes for departure from the overall trend between watershed unit-level AGB density and permafrost. We see that the watershed units furthest below the trend-line (points highlighted in red in Fig.~\ref{ws-map}) correspond to watersheds with the highest proportion of area burned during the 2004 Boundary Fire, suggesting that these areas are still in a state of recovery from fire after a decade. We may expect these two watershed units to continue to accrue AGB until they begin to align with the overall trend between AGB density and permafrost. Fig.~\ref{ws-map} captures the effect of fire history on watershed-unit AGB variability beyond what would be expected due to variation in permafrost coverage at CPCRW.

\section{Comparing \textit{Coregionalization} model and cokriging predictions}
Table \ref{cokriging} presents holdout prediction accuracies and 95\% coverage probabilities for the cokriging interpolations and \emph{Coregionalization} models at BCEF, CPCRW, TNWR and TVSF. The results in Table \ref{cokriging} indicate that cokriging and \emph{Coregionalization} holdout prediction accuracies were similar at all four study sites, although prediction accuracy estimates were slightly better for the \emph{Coregionalization} models at BCEF, CPCRW and TNWR. This is likely due to the restrictions imposed on the fitted variogram models necessary for valid cokriging prediction variances. In contrast, the \emph{Coregionalization} framework estimated using Bayesian inference ensures positive definiteness by construction, meaning that, as long as the covariance functions used for the spatial correlation elements and the $\bK$ variance-covariance matrix are positive definite, the model is estimable. Positive definiteness for the spatial random effect variances and covariances is satisfied through appropriate prior specification, e.g., the $IW$ prior on $\bK$ is a distribution of only positive definite matrices and the spatial decay priors only consider positive values. The increased flexibility in modeling spatial autocorrelation provided by the \emph{Coregionalization} framework estimated using Bayesian inference may be leading to slightly improved prediction over the \texttt{gstat} implementation of cokriging at BCEF, CPCRW and TNWR. As shown in Table \ref{cokriging}, the empirical 95\% coverage probabilities for the \emph{Coregionalization} model holdout uncertainty intervals better match the intended 95\% coverage than the cokriging intervals. The cokriging confidence intervals are far too narrow, leading to low empirical coverage. This is a direct result of the inability of classically estimated cokriging to account for uncertainty associated with the semi- and cross-variogram parameter estimates. These variogram parameters are difficult to estimate with high confidence and methods that fail to account for this uncertainty will produce unreliable coverage intervals. This is a common problem identified for many geostatistical models estimated using frequentist approaches, including kriging and cokriging \citep{sjostedt2003,schelin2010}.   

\begin{table}[h!]
\centering
\caption{Cokriging and \emph{Coregionalization} holdout prediction accuracies and 95\% coverage probabilities. To avoid potential bias introduced due to back-transformation of cokriging predictions and variances, prediction and uncertainty interval coverage is compared on the transformed ($\sqrt{Mg/ha}$) scale.}\label{cokriging}
\begin{tabular}{|r|cc|cc|}\hline
       & \multicolumn{2}{|c|}{CV-RMSE ($\sqrt{Mg/ha}$)} &  \multicolumn{2}{|c|}{95\% Coverage Probability} \\ \hline
       & cokriging & coregionalization & cokriging & coregionalization \\
 BCEF  &   2.23    &       2.14        &   0.739   &        0.947      \\
 CPCRW &   1.75    &       1.63        &   0.651   &        0.946      \\
 TNWR  &   2.04    &       1.96        &   0.722   &        0.963      \\
 TVSF  &   1.94    &       1.96        &   0.620   &        0.945      \\\hline
\end{tabular}
\end{table}

\section{Conclusions and Next Steps}\label{discussion}
The goal of this analysis was to develop and test the performance of a statistical modeling framework that can 1) incorporate partial coverage lidar data and wall-to-wall Landsat products to improve AGB density prediction; and 2) accommodate spatially structured error, thereby allowing for more reliable model-based characterization of uncertainty and improved prediction. Through model comparison we were able to show that a coregionalization framework can effectively be used to couple sampled lidar and field data to improve grid cell-level AGB density prediction accuracy and increase confidence in total AGB estimates. Rigorous model comparison also demonstrated the adverse effects of spatial autocorrelation and how including appropriately specified random effects within a Bayesian hierarchical framework can absorb spatial dependence in the response variable not accounted for by the covariates.

An examination of \emph{Coregionalization} and \emph{Coregionalization$+$Tree Cover} model mapped predictions revealed benefits of including wall-to-wall information to better visualize spatial variability in AGB. We also saw that greater differentiation of AGB median point estimates provided by Landsat-based tree cover data does not necessarily lead to improved grid cell-level predictive performance. It can be the case that simply constructing models to leverage proximate observations may improve prediction more than supplementing auxiliary data, especially when sites are intensively sampled and ancillary information is only weakly related to the response variable. 

We demonstrated the ability to summarize PPDs from the \emph{Coregionalization$+$Tree Cover} model to generate small-area estimates of mean AGB density at CPCRW. Small-area estimates are a challenge for traditional estimation methods \citep{breidenbach2012, goerndt2011}, and have also limited the ability to investigate fine-scale spatial heterogeneity within larger management units such as the TVSF or TNWR. In contrast, our results demonstrate how small-area integrated PPDs can be used to examine correlation with other information about sub-domains while accounting for prediction uncertainty. Relating watershed unit-level AGB density to permafrost and fire history polygon layers uncovered possible drivers of AGB variability among watershed units at CPCRW while explicitly accounting for uncertainty of AGB density predictions. This exercise shows the inferential potential of AGB PPD products generated using Bayesian hierarchical spatial models beyond simply mapping AGB density.

Upcoming satellite lidar missions, e.g., ICESat-2 and GEDI LiDAR, are designed to collect data along orbital transects but not provide complete spatial coverage. Results from this study suggest that it may be possible to implement a coregionalization approach to augment the FIA program's network of permanent sample plots over the contiguous United States with sampled space-based lidar to improve forest inventory estimates. Further, Landsat products can be incorporated to potentially improve point-level prediction accuracy and areal inventory estimates at a national scale. 

The coregionalization framework detailed here is very computationally intensive. After dramatically subsampling the lidar dataset (described in Section \ref{rsdata}), the TVSF \emph{Coregionalization} model took two days to complete a single MCMC chain of 50\,000 iterations on a Linux workstation equipped with an eight-core processor leveraging threaded BLAS and LAPACK C$++$ libraries (\url{www.netlib.org/blas} and \url{www.netlib.org/lapack}). When lidar and/or field observations exceed 10\,000, fitting this class of multivariate spatial models within a Bayesian paradigm becomes intractable on typical desktop computers due to computational difficulties related to inverting massive matrices. Recent advances in $GP$ modeling theory and computation have unearthed new approaches for estimating spatial random effects that circumvent the limitations imposed by the necessity to invert large matrices \citep{datta2016a, datta2016b, finley2017b}. Implementing Nearest-neighbor Gaussian process priors ($NNGP$) in place of traditional $GP$s alleviates the need to invert large covariance matrices, allowing modelers to effectively estimate complex Bayesian spatial models using many remote sensing and field observations on standard workstations. In the future, we will investigate the approximation of $GP$s with $NNGP$s to allow for the inclusion of more field and remote sensing samples within coregionalization frameworks to further improve grid cell-level prediction and areal estimation certainty. These techniques will allow broader applicability of robust modeling techniques to other disciplines, based on the large data volumes typically associated with spatially extensive airborne and satellite remote sensing products. We also plan to extend this class of multivariate spatial models to allow for the simultaneous prediction of multiple forest variables, e.g., tree density, basal area and AGB, while leveraging spatial cross-correlations between all responses. By jointly predicting many forest inventory parameters we will be able to preserve the inherent relationships between them and use those intrinsic correlations to aid spatial prediction and subsequent areal estimation of all included forest inventory parameters.

\section{Acknowledgments}
The research presented in this study was partially supported by NASA's Arctic-Boreal Vulnerability Experiment (ABoVE) and Carbon Monitoring System (CMS) grants (proposal 13-CMS13-0006 funded via solicitation NNH13ZDA001N-CMS and proposal 15-CMS13-0012 funded via solicitation NNH15ZDA001N-CMS). Additional support was provided by the United States Forest Service Pacific Northwest Research Station. Andrew Finley was supported by National Science Foundation (NSF) DMS-1513481, EF-1137309, EF-1241874, and EF-1253225 grants.
\clearpage
\singlespacing
\bibliographystyle{RSOE-references} 
\bibliography{manuscript}
\end{document}